# Lattice dynamical properties and interatomic force constants of transition metal oxide perovskite superlattices

Xiao-Lin Chen,[1,2] Jin-Zhu Zhao,[1,2,3,4*] Philippe Ghosez,[5]

*1. Guangdong Basic Research Center of Excellence for Structure and Fundamental Interactions of Matter, Guangdong Provincial Key Laboratory of Quantum Engineering and Quantum Materials, School of Physics, South China Normal University, Guangzhou 510006, P. R. China;*

*2. Guangdong-Hong Kong Joint Laboratory of Quantum Matter, Frontier Research Institute for Physics, South China Normal University, Guangzhou 510006, P. R. China;*

*3. Center for Computational Science and Engineering, Southern University of Science and Technology, Shenzhen 518055, P. R. China;*

*4. National Laboratory of Solid State Microstructures, Nanjing University, Nanjing 210093, P. R. China.*

*5. Theoretical Materials Physics, Q-MAT, University of Liège, B-4000 Sart Tilman, Belgium.*

## Abstract

In this study, we investigate the construction of efficient model potentials for perovskite heterostructures by transferring relevant dynamic quantities from bulk systems and short-period superlattices. Through first-principles calculations, we provide a comprehensive analysis and comparison of the dynamical properties of different perovskite-based superlattice structures, including $(PbTiO_3)_n/(SrTiO_3)_m$ superlattices, $(ATiO_3)_n/(AO)_2$ superlattices, and $A_{n+1}Ti_nO_{3n+1}$ Ruddlesden-Popper phases, where A can be Sr, Ba, or Pb. We demonstrate that as the thickness of these structures increases, the dynamical properties of the perovskite blocks converge rapidly towards their bulk counterparts. We discuss this convergence in terms of phonon density of states in reciprocal space and inter-atomic force constants in real space. Our findings suggest a straightforward strategy for developing transferable model potentials that accurately capture the dynamics of layered oxide compounds with progressively thicker layers.

Correspondence and requests for materials should be addressed to Jin-Zhu Zhao (zhaojz@m.scnu.edu.cn)

## I. INTRODUCTION

$ABO_3$ perovskite oxides form a technologically important and widely studied family of compounds, which can exhibit various properties including ferroelectricity, piezoelectricity, ferromagnetism, multiferroism, giant magnetoresistance, metal insulator transitions [1–6], etc. Beyond the interest for simple perovskite structures, various types of superlattice structures combining perovskite blocks or alternating the with other structures have also been the subject of intensive research due to the possible emergence of new and exotic properties at the layer interfaces, offering potential interest for technological applications [7–13]. This concerns artificial heterostructures (alike the $SrTiO_3$/$LaAlO_3$ interface [14]) and superlattices (alike $PbTiO_3$/$SrTiO_3$ [9]) as well as naturally-layered structures such as Ruddlesden-Popper [13], Aurivillius or Dion-Jacobson phases [15,16].

During the past years, first-principles calculations based on density functional theory revealed particularly powerful to investigate the physics of perovskite compounds. However, for computational reasons, these calculations are nevertheless typically restricted in practice to relatively small systems (simple perovskite, short-period superlattices and ordered solid solutions) and moreover to zero Kelvin. Substantial efforts have been made to expand the capabilities of theoretical calculations to large periodic systems (for instance tens of thousands atoms per unit cell) and finite temperature behaviors.

Since 1970s, an effective model potential approach relying on first-principles results was developed to overcome these limitations [17–19] and was successfully applied to various simple perovskites [20–28] and solid solutions [29–31]. This approach, based on a projection within a restricted subspace of relevant lattice degrees of freedom has been recently generalized in the form of first-principles effective atomic potentials, to include the whole dynamics [32]. The potential energy surface around the reference phase is described as a Taylor expansion in terms of all the displacements of atoms and strain. Such atomic potential approach not only revealed appropriate to describe simple perovskites like $PbTiO_3$ [32] and $BaTiO_3$ [21], $CaTiO_3$ [24], $PbZrO_3$ [28], $SrTiO_3$ [33], but the bulk parameters were even recently adapted to model large-period $PbTiO_3$/$SrTiO_3$ [34] superlattices. In the meantime, another strategy also provided comparable results, through the mapping of the interatomic interactions via force field interatomic potentials [35,36]. In the early times, the efficient of this approach strongly depended on arbitrary setting and sometimes had to handle complex energy functions [37]. Recently, a promising solution for this challenging emerged [38,39], that was preliminary applied on some representative perovskites [40] and monolayer material [41]. The effective model potential of the system can be obtained through deep neural network-based machine learning on a group of training set based on the DFT results for the potential energy surface [38,39].

Although these theoretical approaches are extremely promising and open exciting perspectives the transfer of bulk parameters to layered structures remains a delicate step, and generalizing the approach to arbitrary systems still require to gain some knowledge on the evolution of the dynamical properties when forming heterostructures.

To some extent, large-period superlattices can be seen as the combination of bulk-like and interfacial regions. In this perspective, there is a reasonable hope that the related effective atomic potential parameters could be derived from first-principles calculations on bulk compounds and short-period superlattices (SSL). This decomposition into bulk and interfacial regions is nevertheless arbitrary; validating a practical approach requires therefore to determine how fast the dynamical properties of layers of increasing thickness converge to their bulk values, in order to identify how thick the interfacial region should be, from a dynamical point of view.

In order to address these questions, we consider here three types of systems and study them form first-principles: $(PbTiO_3)_n/(SrTiO_3)_m$ superlattices, $(ATiO_3)_n/(AO)_1$ Ruddlesden-Popper (RP) compounds and $(ATiO_3)_n/(AO)_2$ artificial (AF) superlattices (with A = Sr, Ba or Pb) as shown in Fig. 1. The relaxed atomic positions and the lattice dynamical properties of these systems were determined. A special emphasis will be placed on the analysis of both the phonon density of states (*DOS*) and the inter-atomic force constants (*IFCs*), and to the comparison with their counterpart in the related bulk compounds. Through our discussion, we will see that the dynamical properties of the superlattices converge very quickly to the bulk values when going away from the interface, suggesting a simple and straightforward approach for constructing atomic potentials for arbitrary superlattices.

## II. METHODS

### A. Systems of Interest

In what follows, we make use of a supercell approach and periodic boundary conditions. By convention, in all superlattices, the plane parallel to the interface will correspond to the $(x, y)$ plane, while the perpendicular directions will be referred to as the $z$ axis. The layers and bonds are indexed starting from the interface according to the notations defined in Fig. 1.

$(PbTiO_3)_n/(SrTiO_3)_m$ superlattices (sometimes labeled as P$n$S$m$ in the following) are made from the periodic repetition of n layers of $PbTiO_3$ alternating with m layers of $SrTiO_3$. In this work, we consider $(n, m)$ equal to (1, 1), (1, 3) and (3, 1). For these superlattices, we consider an epitaxial growth on a $SrTiO_3$ substrate so that the in-plane lattice constant is constrained to the value of bulk $SrTiO_3$ obtained from first-principles. As the references, we employ bulk $SrTiO_3$ and strained bulk $PbTiO_3$ with the in-plane lattice constants fixed to the value of cubic $SrTiO_3$. We further restrict ourselves to the high-symmetry *P4/mmm* phase, relaxing together the $c$ axis (along the $z$ direction) and the internal atomic degrees of freedom.

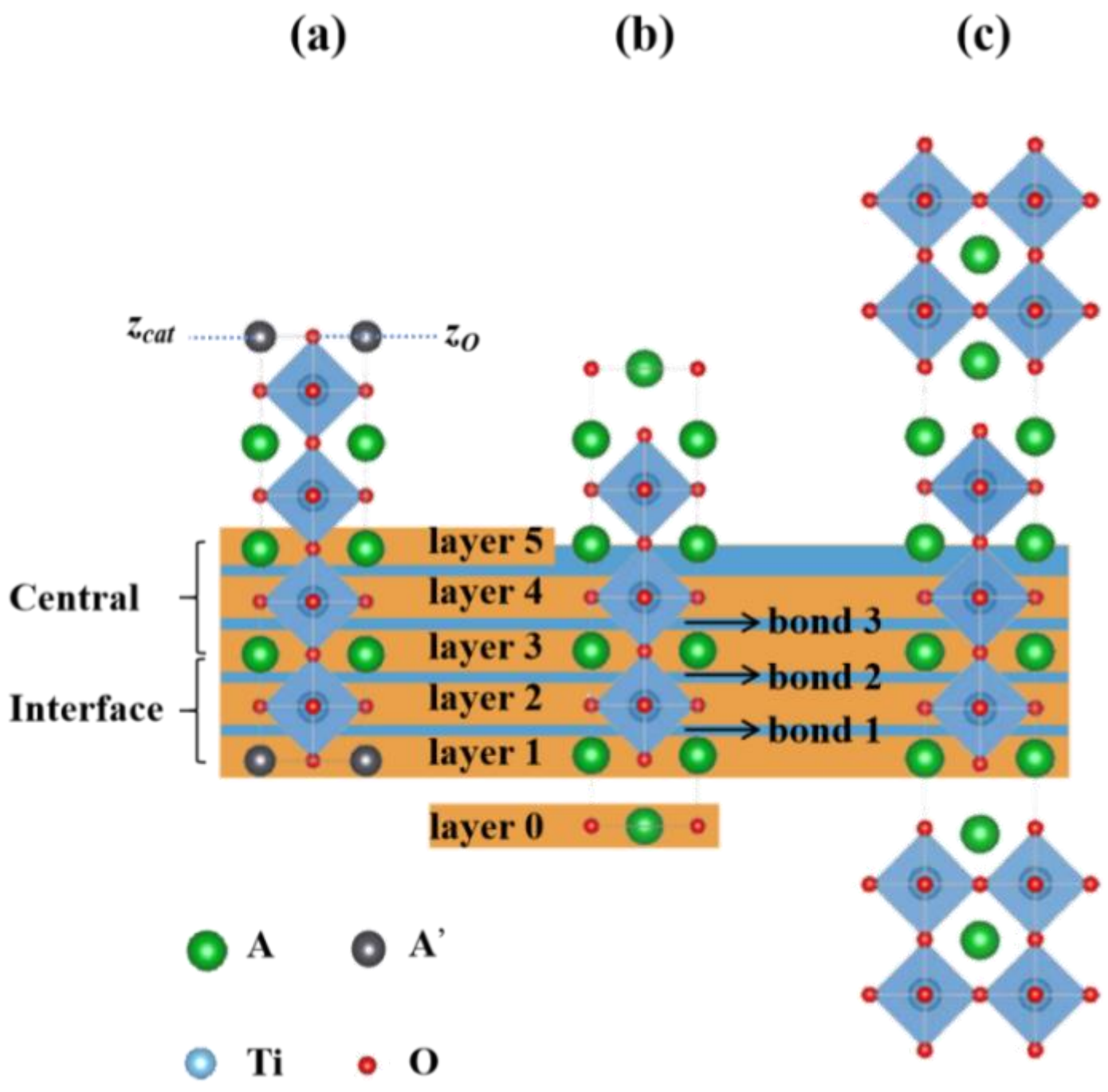

**Figure 1.** Systems of interest and labelling of each layer (orange color) and bond (blue color). Here, A refers to Ba, Sr and Pb. To further distinguish different cations in $(PTO)_n/(STO)_m$ superlattices, A' refers to Sr and Pb. (a) $(PTO)_n/(STO)_m$ superlattices in the case of $n$=1, $m$=3; (b) artificial $(ATiO_3)_n/(AO)_m$ superlattices in the case of $n$=3, $m$=2; (c) Ruddlesden-Popper $(ATiO_3)_n/(AO)_m$ superlattices in the case of $n$=3, $m$=1. The atomic rumpling of each layer and related average plane are defined in (a).

The other two series of superlattices have the generic formula $(ATiO_3)_n/(AO)_m$ with $m$=1 for the RP phases and $m$=2 for the AF superlattices. In this work, we will consider explicitly A=Ba, Sr, Pb and $n$=1, 2 and 3. As shown in Fig. 1, the structure of these two types of superlattices can be considered as a perovskite $ABO_3$ layer epitaxially grown on a rocksalt AO block. In the RP compounds, the perovskite unit cells are shifted along *x*-*y* direction by (1/2,1/2) lattice constant after each AO interface so that the unit cell along *z* is doubled. For these systems, we also focus on the highly-symmetric *P4*/*mmm* phase but no epitaxial constraint is imposed and the lattice parameters and atomic degrees of freedom were fully relaxed together.

**B. Technical Details**

Our first-principles calculations were performed in the framework of density functional theory (DFT) as implemented in the VASP package [42,43]. For the three systems, the calculations were done within the PBEsol functional under the generalized gradient approximation (GGA). We used the projected augmented wave (PAW) method to describe the interaction energy between valence electrons and ions. The following states were treated as valence states: the *5s*, *5p* and *6s* states of Ba, the *4s*, *4p* and *5s* states of Sr, the *5d*, *6s* and the *6p* states of Pb, the *3s*, *3p*, *3d* and *4s* states of Ti and the *2s* and *2p* states of O. The electronic wave functions were expanded in a plane-wave basis; in all cases, we used an energy cutoff of 600 eV, which guarantees a very good convergence of the results. Integrals over the Brillouin zone were achieved using gamma-centered meshes of k-points equivalent (in terms of density) to a 6×6×6 sampling for the simple cubic perovskite. The vibrational properties were calculated using density functional perturbation theory (DFPT) [44]. The Berry Phase Method [45] is used to calculate electric polarization.

The lattice constants and atomic positions were relaxed while imposing the *P4/mmm* symmetry (space group No. 123) for the superlattice structures and *Pm-3m* cubic symmetry (space group No. 221) for related bulk perovskites. For the cubic phase of bulk $ABO_3$, a *n*-layers supercell $(ABO_3)_n$ was used in order to compare the results at exactly the same level of convergence. We elaborate further on the value of *n* and convergence aspects in the Supplementary Materials (SM).

## III. RESULTS AND DISCUSSIONS

### A. $(PTO)_n/(STO)_m$

#### *1. Structural convergence*

Firstly, we investigate the $(PTO)_n/(STO)_m$ superlattices in the *P4*/mmm phase. The presence of interfaces introduces atomic distortions to the structure compared to ideal perovskites. To analyze the effect of the interface on atomic structures, we focus on three quantities for characterizing the structural distortions. Firstly, we consider the lattice constant along the *z* component, as the in-plane values were fixed. In Fig. 1, we define the lattice constant of the interface region ($c_i$) and the central region ($c_c$) as the distance between the average plane coordinates for each layer, which we define as $z_{aver}=(z_{cat}+z_O)/2$. Secondly, we report the rumpling of each atomic layer, defined as $R_{rump}=(z_{cat}-z_O)/2$, and thirdly, the Ti-O bond length. The calculated values of these three quantities are presented in Table I. We observe that the values in the interface region (P1S3 and P3S1, defined as large-period superlattices) are close to those of the related short-period superlattice (P1S1), while the lattice constant and the values in the central region almost recover the corresponding bulk ones. For instance, the lattice constant and the Ti-O bond length converge with an error below 0.6% (see Table I). Furthermore, the rumpling primarily occurs in the interface region and quickly diminishes towards the central region. Layer 2 ($TiO_2$ type) exhibits the maximum rumpling amplitude, approximately 0.014 Å in each system. From layer 3 to layer 5, the rumpling of each layer becomes negligible, indicating a rapid convergence of atomic structures to bulk values. The relaxation of the interface does not propagate beyond one perovskite layer (4 Å) from the interface.

**Table I.** Lattice size and thicknesses of central layer $c_c$ and interface layer $c_i$ (in Å) in $(PTO)_n/(STO)_m$ (P1S1, P1S3 and P3S1), compared with related values in bulk cubic $SrTiO_3$ and bulk tetragonal $PbTiO_3$. Value of rumpling for each layer (in Å) in $(PTO)_n/(STO)_m$. The length of Ti-O bond along $z$ in $(PTO)_n/(STO)_m$ compared with related values in bulk cubic $SrTiO_3$ and bulk tetragonal $PbTiO_3$. Each quantities are labeled in Fig. 1.

| | Lattice size | | Atomic rumpling | | | | | Ti-O bond length | | | |
|---|---|---|---|---|---|---|---|---|---|---|---|
| | $c_i$ | $c_c$ | layer 1 | layer 2 | layer 3 | layer 4 | layer 5 | bond 1 | bond 2 | bond 3 | bond 4 |
| P1S1 | 3.920 | 3.900 | 0 | -0.014 | 0.001 | 0 | 0 | 1.970 | 1.948 | 1.950 | 1.952 |
| P3S1 | 3.922 | 3.943 | 0 | 0.014 | 0.001 | 0 | 0 | 1.949 | 1.974 | 1.971 | 1.971 |
| P1S1 | 3.923 | | 0 | 0.014 | 0 | — | — | 1.948 | 1.976 | — | — |
| $PbTiO_3$ | 3.941 | | 0 | | — | — | — | 1.971 | | — | — |
| $SrTiO_3$ | 3.899 | | 0 | | — | — | — | 1.950 | | — | — |

*2. Phonon Density of states*

Since the real-space structure of the superlattices converges to the bulk case quickly, we expect that the dynamics of the superlattices may also recover the bulk behavior in the central layer due to its similar atomic environment. Following this idea, we project the *DOS* onto each atomic layer. Thus, the dynamical contribution of the entire superlattice is divided into two parts: the central layer and the interface layer, as shown in Fig. 2, where the negative values of vibration frequencies correspond to unstable soft modes.

Significant differences are observed between the *DOS* of the central layer and the interface layer. For instance, in the frequency range from 0 THz to 5 THz, the interface layer exhibits two main peaks. The first peak, around 1.1 THz, corresponds to the contribution of Pb atoms, while the second peak, around 3.2 THz, corresponds to the contribution of Sr atoms. In the central layer of P3S1 and P1S3 superlattices, there is only one peak in each case. Comparing with the interface region, the peak corresponding to the Pb atom in the central layer of P3S1 has a higher frequency of 1.1 THz, whereas the peak corresponding to the Sr atom in the central layer of P1S3 has a lower frequency of 3.1 THz.

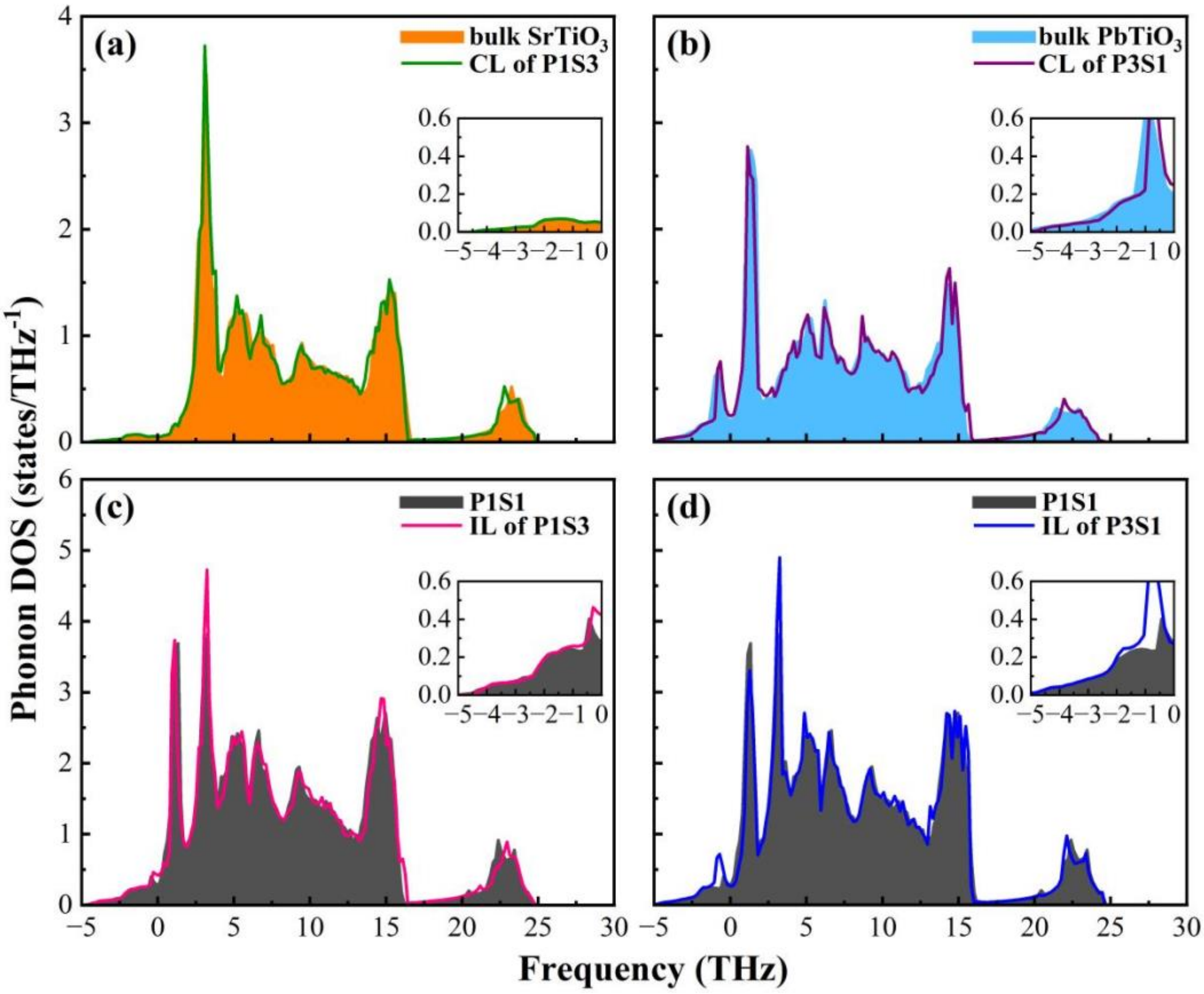

**Figure 2.** Phonon density of states (*DOS*) of interface and central layers in P1S3 and P3S1 cases, and the phonon *DOS* for bulk. The region of negative frequencies is zoomed as insets. (a) bulk cubic $SrTiO_3$ (orange line) and central layer in P1S3 (green line); (b) bulk tetragonal $PbTiO_3$ (light blue line) and central layer in P3S1 (purple line); (c) P1S1 case (dark gray line) and interface layer in P1S3 (pink line); (d) P1S1 case (dark gray line) and interface layer in P3S1 (blue line).

As shown in Fig. 2, all these features of the layer-projected phonon *DOS* for the superlattices are nicely reproduced by the related bulk cases and the short-period P1S1 case, respectively. Thus, we observe that the phonon *DOS* of P1S1 can captures the contribution of the interface region in both P1S3 and P3S1 superlattices, while the phonon *DOS* of bulk $PbTiO_3$ and $SrTiO_3$ recovers the contributions of the central region in P3S1 and P1S3 superlattices, respectively. Overall, this recovery is of high quality, not only in terms of shape but also for both the frequency and amplitude of each peak. It is interesting to note that all these observations are achieved even when the interface region is only one perovskite layer thick.

*3. Inter-atomic Force Constants*

According to the aforementioned results, the nice convergence behaviors of lattice dynamics support our hypothesis that it would be efficient to decompose the studied superlattices into a central region and an extremely thin interface region. This also suggests that in the approach of effective atomic potential for large-period superlattices, the harmonic term of energy expansion can be directly obtained from the first-principles results of the bulk materials and SSL. Therefore, the inter-atomic force constants (*IFCs*) in real space, which serve as transferable physical quantities and key inputs for constructing effective atomic potential, deserve further investigation as they are the origin of phonon *DOS* in the studied materials. In principle, the amplitude of *IFCs* decreases with increasing inter-atomic distance, and only the atoms in a relatively small region around a certain atom contribute significantly to its dynamic behavior. In other words, the *IFCs* from a finite number of atoms near the reference atom need to be considered. In line with these considerations, we will now discuss how to determine the appropriate distance for such spatial truncation.

In this analysis, we select the representative Ti atom as the reference and artificially set the spatial truncation distance to 7 Å. The absolute values of the *IFCs* for atoms within the truncation distance, excluding those with *IFC* $< 0.1$ eV/Å$^2$ and some equivalent atoms, are reported in Fig. 3 and Fig. 4. Furthermore, we compare these values with those of the corresponding bulk and SSL cases, as shown in Fig. 3 and Fig. 4, respectively. Generally, the values of *IFCs* decrease with increasing inter-atomic distance. However, for instance, the maximum *IFC* values is observed between the Ti atom and its nearest neighbor Ti atom, which is approximately 15.6 eV/Å$^2$, indicating the significance of the Ti-Ti interaction. When the inter-atomic distance exceeds 4 Å, the amplitude of each *IFC* values becomes smaller than 2.3 eV/Å$^2$. In both the P1S3 and P3S1 cases, although the phonon *DOS* of the central layer and the interface layer exhibit significant differences, the *IFC* values that affect the Ti atoms in these two regions are similar, with relatively small differences between them.

To further illustrate these differences, we report the difference values between the *IFCs* on Ti atoms in the central layer and the interface layer of the superlattice, compared with their counterparts in the corresponding bulk cases. It is observed that the *IFC* values for Ti in the central layer of the P1S3 superlattice almost match those in bulk $SrTiO_3$, as the difference values reported in Fig. 3 (c) are very small. Similarly, in Fig. 4 (c), we can see that the *IFC* values for Ti in the central layer of the P3S1 superlattice closely resemble those in bulk tetragonal $PbTiO_3$. Likewise, the difference values reported in Fig. 3 (f) and Fig. 4 (f) also indicate that the *IFC* values for Ti in the interface region nicely recover the ones in the P1S1 case. In contrast, there are significant differences not only between the Ti atoms in the central region and those in the P1S1 case, but also between the Ti atoms in the interface and those in the bulk region.

Here, two points need to be highlighted. First, within the spatial truncation, the difference in *IFC* values of all atoms is less than 0.9 eV/Å², as it is shown in Fig. 3 (c), (f) and Fig. 4 (c), (f). This indicates that the inter-atomic constants in the interface layer and central layer of large-period superlattices would recover to bulk materials and related short-period superlattices. Secondly, as illustrated in Fig. 3 (b) and (c), the main difference values arise from the region where the inter-atomic distance is below 4 Å. This suggests that the differences in phonon *DOS* shown in Fig. 2 are mainly determined by the *IFCs* within one perovskite cell, such as the interactions between the reference Ti atom and its nearest neighbor A-site atoms, O atoms, and other Ti atoms. These features hold for both P1S3 and P3S1 superlattices, which suggests that the *IFC* values in the gray region of Fig. 3 and Fig. 4 can be considered negligible with regards to the lattice dynamics. In other words, the spatial truncation distance for selecting transferable *IFC* values can be very small, i.e. less than 4 Å.

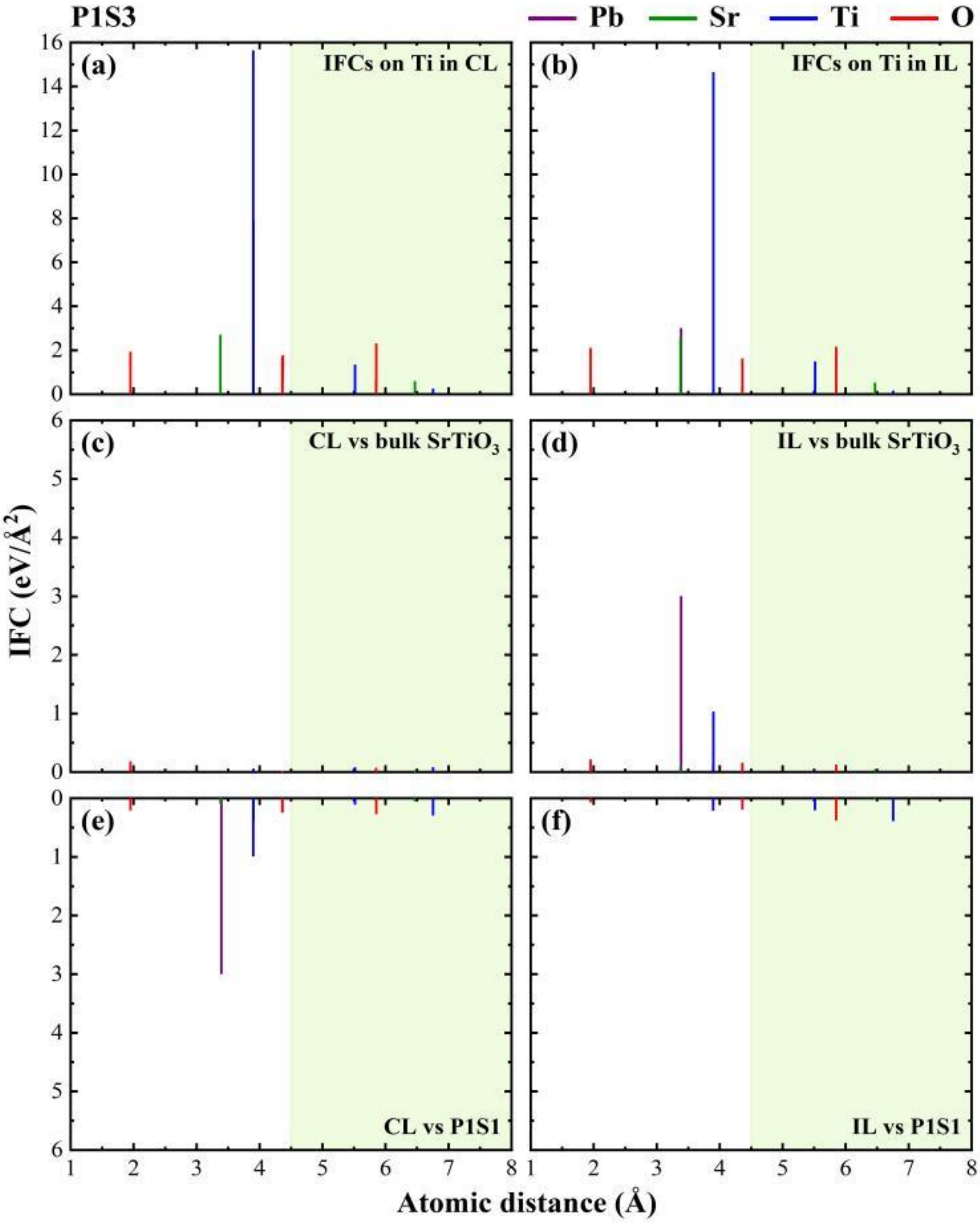

**Figure 3.** Absolute value of *IFCs* difference affect on Ti atoms in selected systems: (a) the central layer of P1S3 and (b) the interface layer of P1S3. Absolute difference between values of *IFCs* affect on Ti atoms in different selected systems: the difference between *IFCs*' for Ti atom. (c) in bulk $SrTiO_3$ and the ones in central layer of P1S3; (d) in bulk $SrTiO_3$ and the ones in interface layer of P1S3; (e) in SSL P1S1 and the ones in central layer of P1S3; (f) in SSL P1S1 and the ones in interface layer of P1S3.

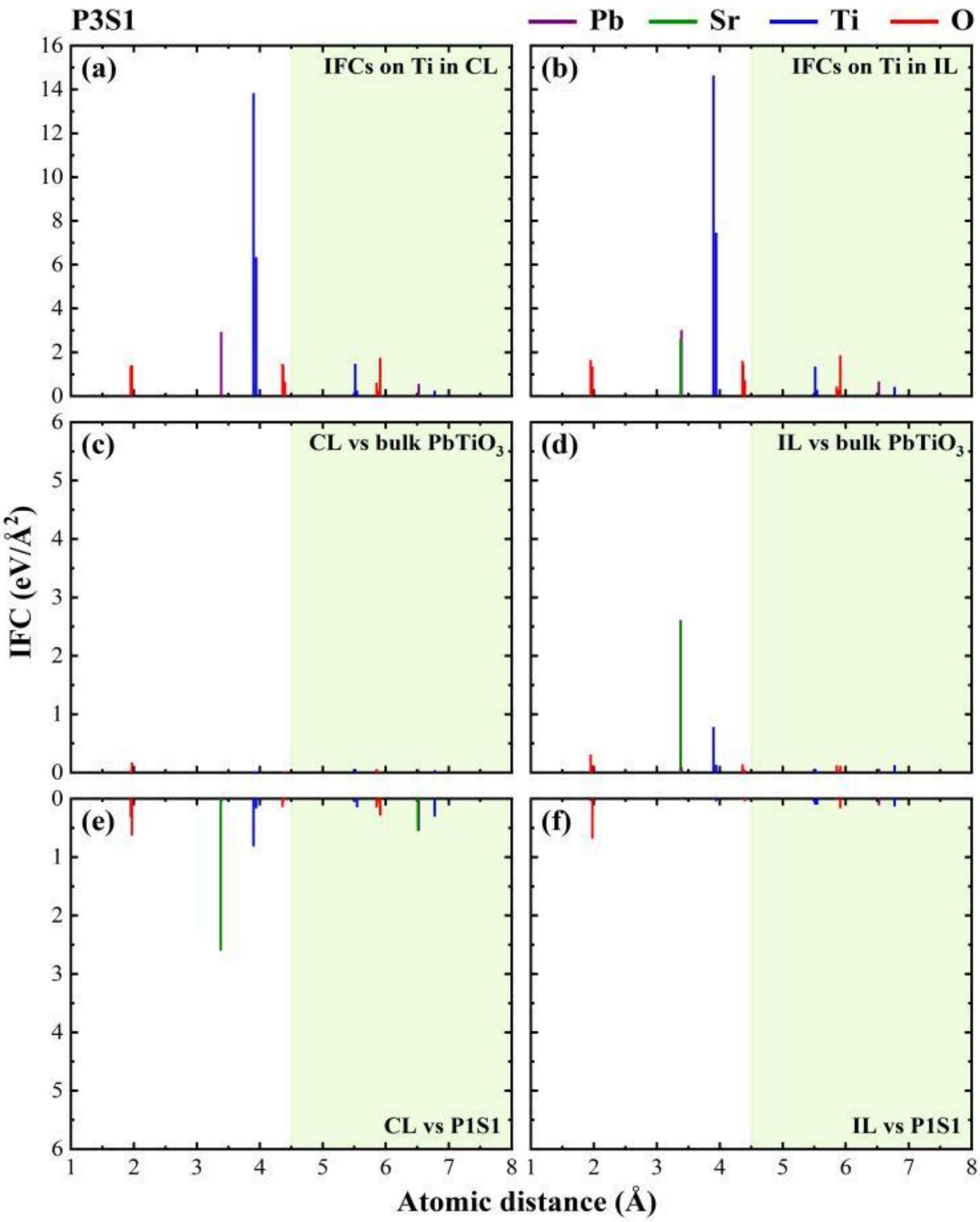

**Figure 4.** Absolute value of *IFCs* difference affect on Ti atoms in difference selected systems: (a) the central layer of P3S1, (b) the interface layer of P3S1. And the absolute difference between values of *IFCs* affect on Ti atoms in different selected systems: the difference between *IFCs*' for Ti atom. (c) in bulk tetragonal $PbTiO_3$ and the ones in central layer of P3S1; (d) in tetragonal bulk $PbTiO_3$ and the ones in interface layer of P3S1; (e) in SSL P1S1 and the ones in central layer of P3S1; (f) in SSL P1S1 and the ones in interface layer of P3S1.

**B. $(ATiO_3)_n/(AO)_m$**

It is worth mentioning that, as previously reported in $PbTiO_3/SrTiO_3$, directly using the average values between the two types of bulk compounds that form the interface, can provide an efficient model. Indeed, this approximation works well when the interfacial distortions are limited with respect to the bulk cases. However, this approach cannot be simply extended to more general cases where the interface is formed by introducing different types of structures, such as interfaces with inserted rocksalt layers in naturally formed RP superlattices. Therefore, it is desirable to extend our method by using the values of *IFCs* in the corresponding short-period superlattices to replace the ones in the interface region of thick superlattices. With this concern in mind, we further examine our assumption in the superlattices, both RP and AF types, composed of perovskite and rocksalt structures, in this section. In the main text, we will present the results of representative systems based on Ba, while the results of systems based on Sr and Pb will be shown in the Supplementary Material (SM).

*1. Structural convergence*

All the structures were relaxed without any epitaxial strain. As reported in Table II, the behavior of the structural convergence is very similar to that observed in $(PTO)_n/(STO)_m$ systems as n increases. Consequently, lattice distortions, in comparison to the ideal cubic perovskite, primarily appear in the interface region and rapidly diminish towards the central region, reaching the bulk value. This convergence is characterized by various structural quantities, including the in-plane lattice parameter(a), the thicknesses of the central layer ($c_c$) and interface layer ($c_i$), the atomic rumpling ($R_{rump}$), and the Ti-O bond lengths along the z-direction. The formulas for calculating these quantities are equivalent to those used in $(PTO)_n/(STO)_m$ systems. Whether in AF or RP structures, the thickness of the central layer in large-period superlattices ($n$=3) closely approximates the value of bulk $BaTiO_3$. Similarly, the values of the interface layer nicely reproduce those observed in corresponding short-period cases ($n$=2).

**Table II.** In-plane lattice constants (a), thicknesses of interface perovskite layer ($c_i$) and central perovskite layer ($c_c$), rumpling in the interface perovskite layer, which is defined as the displacement along $z$ between cations and oxygen atoms $R_{rump}=(z_{cat}-z_O)/2$, Ti-O bond length, under zero epitaxial strain of Ba based superlattices for AF and RP types (all distances in Å)

| | | Lattice size | | | Atomic rumpling | | | Ti-O bond length | | |
|---|---|---|---|---|---|---|---|---|---|---|
| | | a | $c_i$ | $c_c$ | layer 1 | layer 2 | layer 3 | bond 1 | bond 2 | bond 3 |
| AF | n=1 | 3.981 | 3.922 | — | 0.025 | — | — | 1.986 | — | — |
| | n=2 | 3.986 | 3.945 | — | 0.031 | 0.006 | — | 1.955 | 2.023 | — |
| | n=3 | 3.987 | 3.954 | 3.973 | 0.032 | 0.008 | 0.006 | 1.949 | 2.031 | 1.992 |
| RP | n=1 | 3.999 | 3.900 | — | 0.025 | — | — | 1.975 | — | — |
| | n=2 | 3.996 | 3.937 | — | 0.031 | 0.006 | — | 1.943 | 2.025 | — |
| | n=3 | 3.955 | 3.944 | 3.967 | 0.032 | 0.008 | 0.006 | 1.938 | 2.032 | 1.990 |
| $BaTiO_3$ | | | 3.988 | | | 0 | | | 1.994 | — |

*2. Phonon Density of States*

In Fig. 5, we present the results for representative cases of Ba-based AF and RP superlattices. The insets provide a zoomed-in view of the region of negative frequencies. The results for Sr and Pb series superlattices are reported in the SM. It is evident that the phonon density of states (*DOS*) for the central perovskite layer (in the case of $n$=3) differs from that of the interface perovskite layer, particularly in the negative frequency range. For the AF structure ($n$=3), as depicted in Fig. 5 (a), the unstable modes in the central layer, corresponding to the ferroelectric instability, form a peak in the range of -7.5 THz to -2.5 THz, which is consistent with values for bulk $BaTiO_3$. However, as shown in Fig. 5 (c), this peak weakens in amplitude and its frequency shifts to a higher range (-5 THz to 0 THz) for the interface layer, which is accurately reproduced by the corresponding short-period superlattice (n=2). Similar observations are made for the RP compounds, as seen in Fig. 5 (b) and (d). It is worth to notice that these unstable modes only correspond to the in-plane polar motion since the long-range Ti-O correlation along out-of-plane direction is broken due to the interface. Detailed results of the eigenvectors and the polarization driven by these modes are reported in the Sec. II. A of the SM.

These findings confirm the following: i) the ferroelectric instability of cubic $BaTiO_3$ is attenuated in AF and RP superlattices for $n$=3; ii) the central perovskite layer predominantly contributes to the ferroelectric instability, which is suppressed in the interface perovskite layer; iii) the phonon *DOS* contribution of the interface region can be estimated from the *DOS* of the short-period system ($n$=2) in both AF and RP systems.

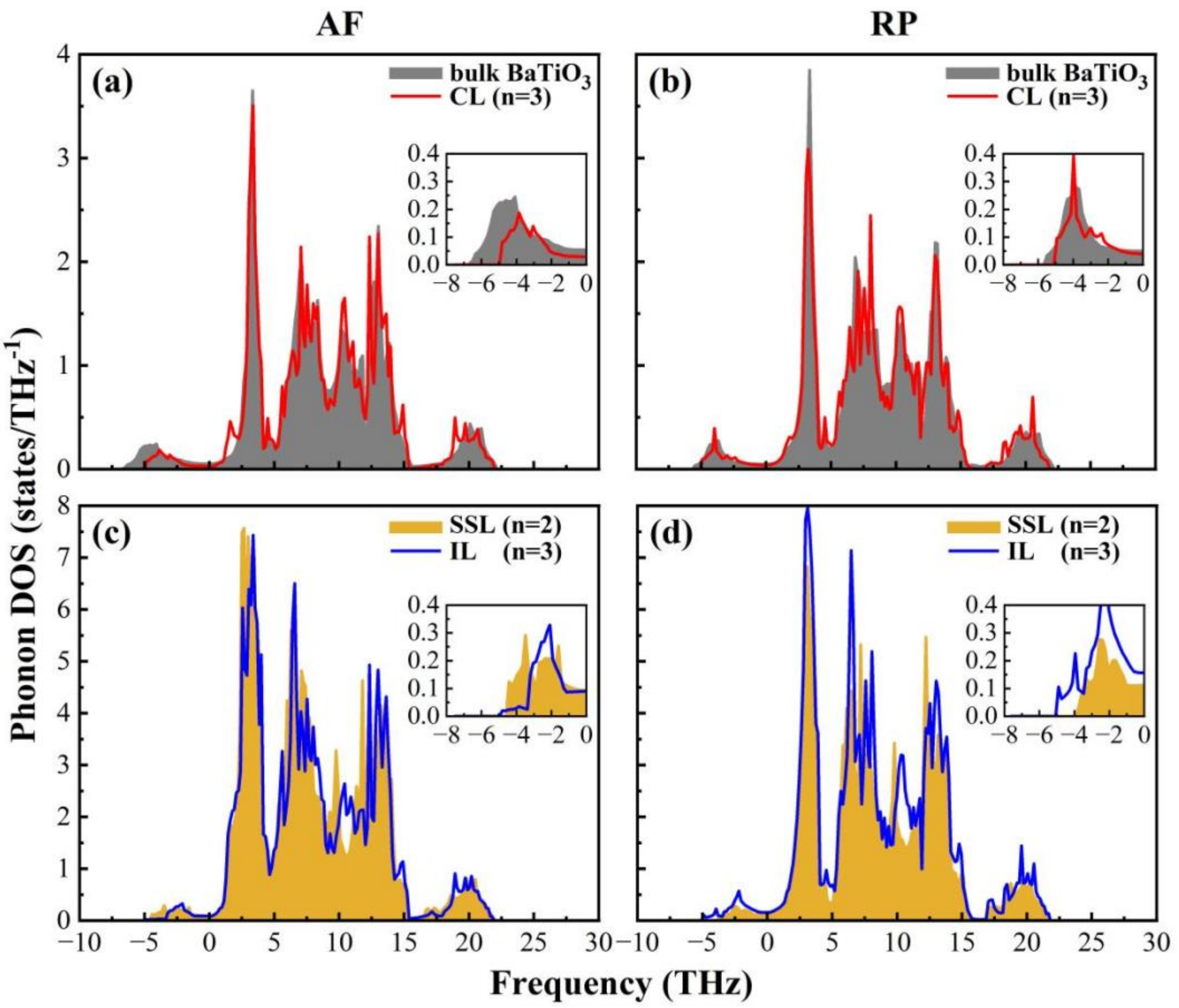

**Figure 5.** Phonon density of states (*DOS*) of Ba based superlattices in the case of AF and RP types. The region of negative frequencies is zoomed as the inset. (a) and (b), the light grey color part refers the phonon *DOS* of the bulk BaTiO3, the red line indicates the phonon *DOS* contribution of the central perovskite layer in *n*=3 superlattices; (c) and (d), the blue line indicates the phonon *DOS* contribution of the interface perovskite layer in *n*=3 superlattices, the dark yellow color refers the contribution from the short-period superlattices in case of *n*=2.

### 3. *Inter-atomic Force Constants*

In this section, we present the *IFCs* values that affect different Ti atoms in the central perovskite layer and interface perovskite layer of Ba-based AF and RP superlattices (n=3). We also illustrate the differences in *IFC* values between these Ti atoms and their counterparts in the bulk system and short-period superlattice (n=2).

As depicted in Fig. 6 (a), (b), and Fig. 7 (a), (b), the absolute values of the *IFCs* for Ti atoms are highly similar between the AF and RP superlattice cases. However, there are noticeable differences in the *IFCs* between the Ti atom in the interface layer and its first neighbor O atom along the z-direction compared to the central layer. This difference can be attributed to the structural distortion in the interface layer. Fig. 6 (c) and Fig. 7 (c) demonstrate that the distorted Ti-O bond length along the z-direction slightly alters the corresponding *IFCs* value, when compared to the bulk $BaTiO_3$. However, the *IFC* values of the central layer in both AF and RP superlattices (n=3) and the *IFC* values of the bulk $BaTiO_3$ are close. Similar observations can be made from Fig. 6 (f) and Fig. 7 (f).

The convergence of inter-atomic force constants (*IFCs*) in Ba-based AF and RP superlattices exhibits a behavior similar to that observed in $(PTO)_n/(STO)_m$ systems. Therefore, the *IFC* values in the central and interface layers can be accurately reproduced using the bulk values and those of the short-period superlattice (SSL) case (*n*=2), respectively. These findings further support our previous remark (mentioned above) that when extracting *IFC* values from DFT data, it may be efficient to limit the spatial truncation to within one perovskite cell. For instance, in Fig. 6 (f) and Fig. 7 (f), the difference in values with the O atom at around 6 Å is relatively larger compared to the rest, and it is also comparable to the differences observed in Fig. 6 (d) and Fig. 7 (d). However, it appears that this discrepancy does not significantly impact the corresponding dynamics. This suggests that the lattice dynamical behavior is predominantly governed by the *IFCs* in a small spatial region, while the values in the light green region of Fig. 6 and Fig. 7 can be considered negligible.

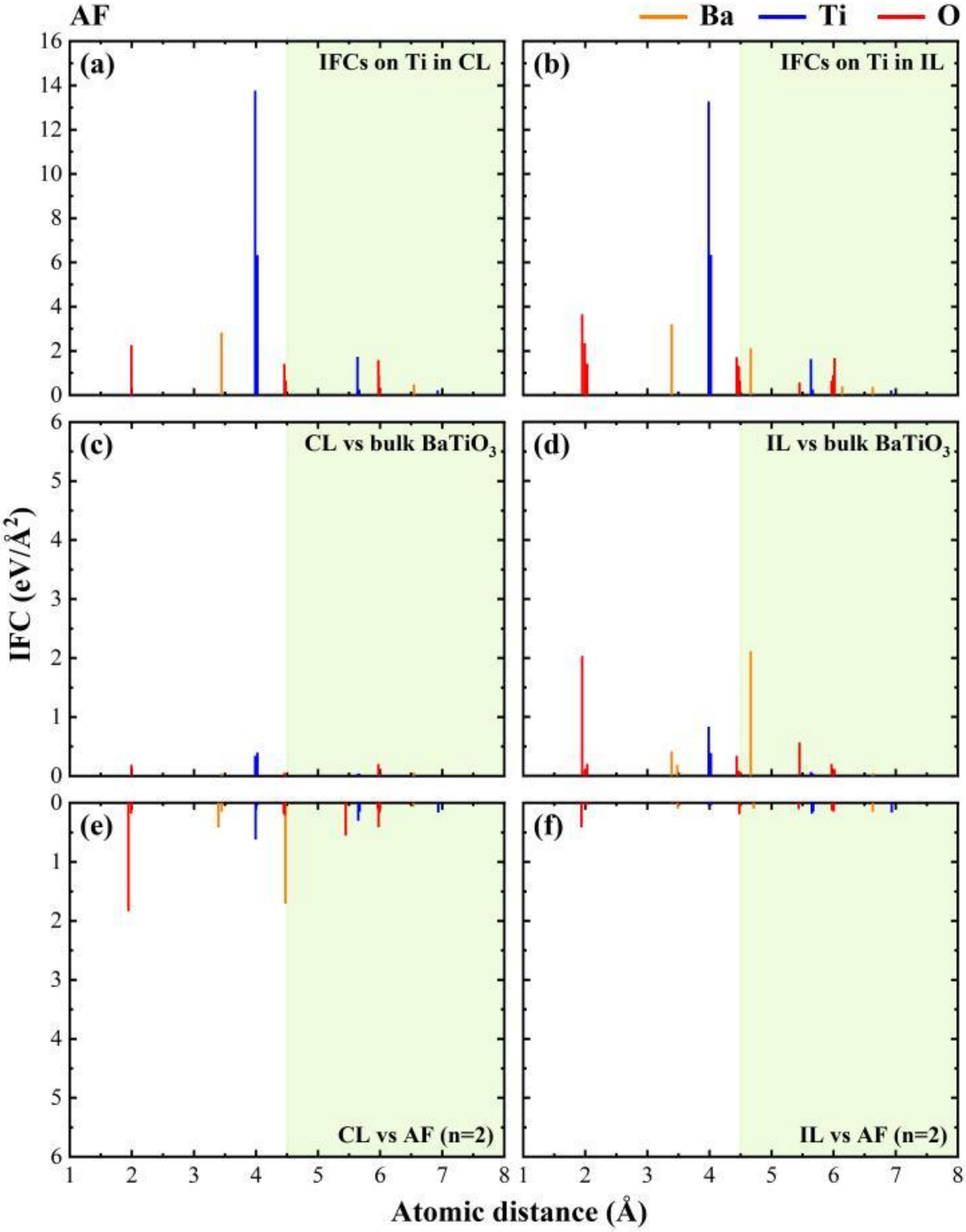

**Figure 6.** Absolute difference between values of inter-atomic constants (*IFCs*) affect on Ti atoms in different selected systems: (a) the central layer of AF superlattice (*n*=3), (b) the interface layer of AF superlattice (*n*=3). Difference between *IFCs*' for Ti atom (c) in bulk $BaTiO_3$ and the ones in central layer of AF superlattice (*n*=3); (d) in bulk $BaTiO_3$ and the ones in interface layer of AF superlattice (*n*=3); (e) in short-period AF superlattice (*n*=2) and the ones in central layer AF superlattice (*n*=3); (f) in short-period AF superlattice (*n*=2) and the ones in interface layer AF superlattice (*n*=3).

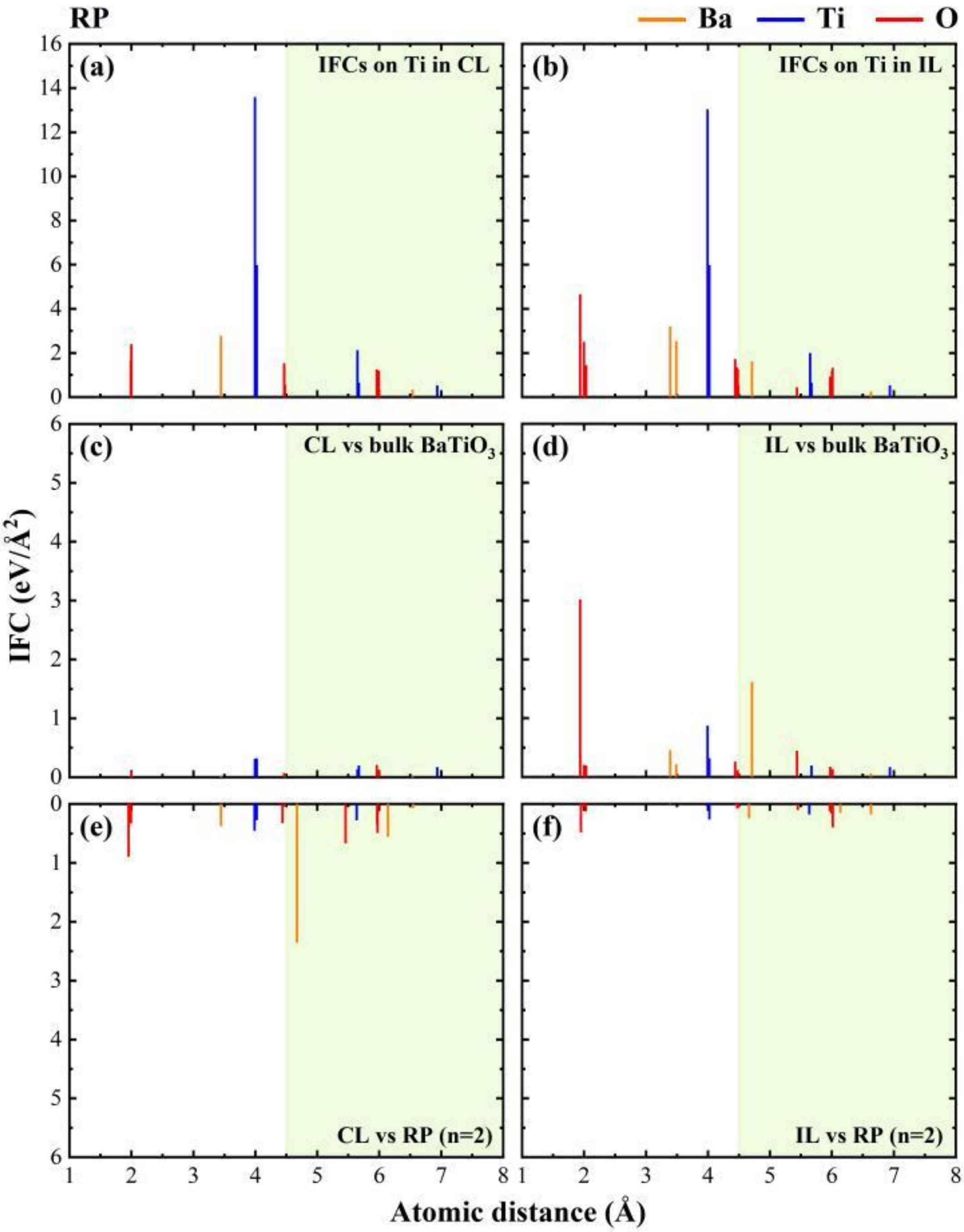

**Figure 7.** Absolute difference between values of inter-atomic constants (*IFCs*) affect on Ti atoms in different selected systems: (a) the central layer of RP superlattice (*n*=3), (b) the interface layer of RP superlattice (*n*=3). Disfference between *IFCs*' for Ti atom (c) in bulk $BaTiO_3$ and the ones in central layer of RP superlattice (*n*=3); (d) in bulk $BaTiO_3$ and the ones in interface layer of RP superlattice (*n*=3); (e) in short-period RP superlattice (*n*=2) and the ones in central layer RP superlattice (*n*=3); (f) in short-period RP superlattice (*n*=2) and the ones in interface layer RP superlattice (*n*=3).

## IV. ARTIFICIALLY CONSTRUCTED PHONON

Based on the results shown above, we preliminarily proposed that the lattice dynamic properties of large-periodic superlattice can be estimated from related bulk and short-period samples. To verify this approach, we first check the truncation value of the *IFCs* showing that the ones can be ignored when the value of which below 0.1 eV/Å$^2$ corresponding the distance of one perovskite layer in real space (see the Sec. II C of SM).

Furthermore, we reconstruct the *IFCs* matrix for the large-period superlattice P1S3 to produce the phonon dispersion curves and related density of states. Along with our proposal, the *IFCs* of central layer in P1S3 are replaced by the values of $SrTiO_3$ and the ones of interface layer are replaced by the values calculated from P1S1 case. For *IFCs* corresponding two atoms that across the central and interface layers, we use the averaged *IFC* values of bulk $SrTiO_3$ and P1S1 superlattice. As reported in Fig. 8, the phonon dispersion curves and phonon *DOS* of P1S3 superlattice generated by the artificially constructed (A.C.) *IFCs* matrix mentioned above are in nice agreement with the ones obtained directly from first-principles. Most of the phonon branches are nicely recovered especially those unstable ones at high symmetric points of Brillouin Zone. Only few errors appear at the lines between high symmetric points. These differences are negligible and suppressed when integrated in the *DOS* where the main features of each peaks, position and amplitude, are in good agreement.

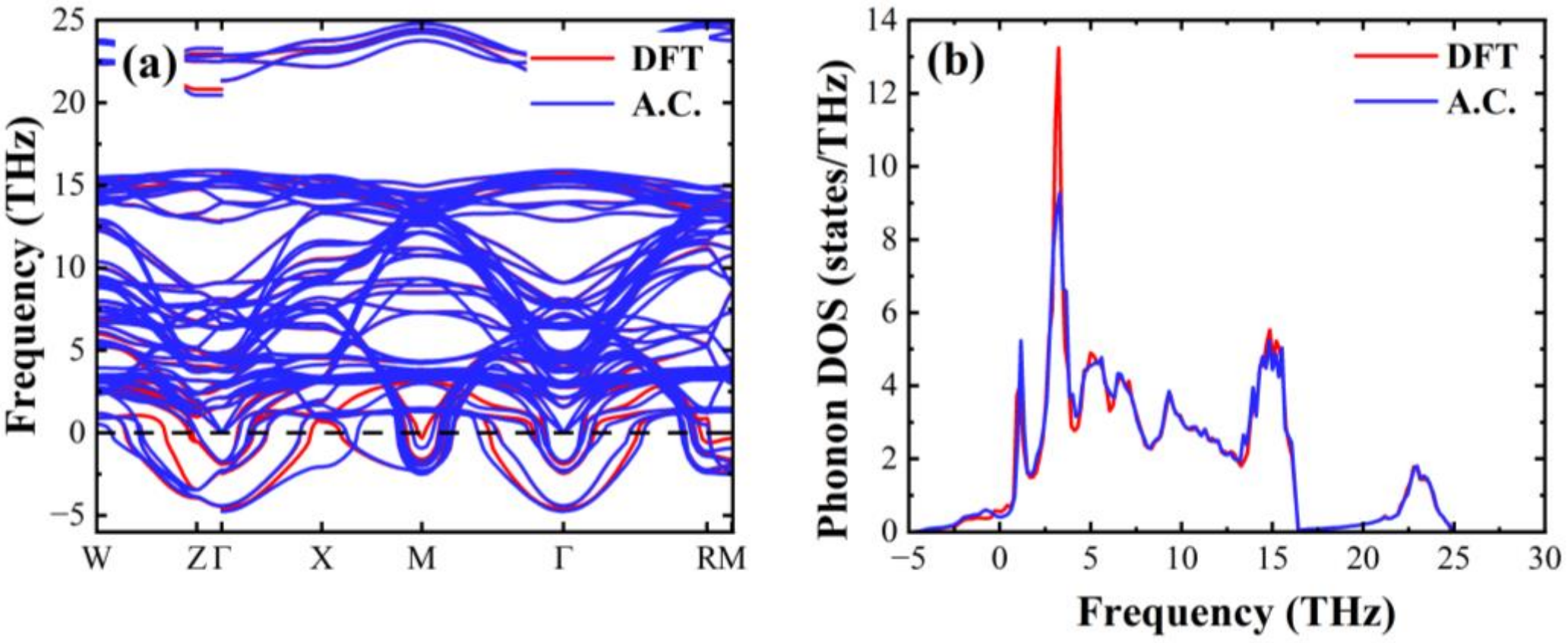

**Figure 8.** The phonon dispersion curves and phonon density of states of P1S3 superlattice. (a) phonon dispersion curves; (b) phonon density of states. The rad line indicates the results of DFT directly, the blue line indicates the results of artificially constructed (A.C.) *IFCs* matrix by bulk $SrTiO_3$ and P1S1 superlattice.

## V. CONCLUSIONS

From the results of structure relaxation for $(PTO)_n/(STO)_m$ and $(ABO_3)_n/(AO)_m$ series superlattices, it is observed that the structure distortion, compared to the high symmetry reference structure, mainly occurs at the interface region and quickly diminishes towards the central part. This convergence in structure is supported by the values of lattice size, atomic rumpling, and representative Ti-O bond length. Similar distortions are found in interfaces of the same type, indicating that the interface structure of large-period superlattices is comparable to that of short-period ones.

This rapid convergence ensures the convergence of layer-dependent dynamical behaviors. The phonon *DOS*, projected in reciprocal space, reveals differences in phonon contributions between the interface and central perovskite layers, especially in the negative frequencies' region (unstable modes). The phonon behavior in this region not only distinguishes different layers but also demonstrates that similar atomic structures yield similar phonon *DOS* contributions. According to our results, the dynamics of the central perovskite layer converge well towards the bulk cubic dynamics, which is only one perovskite layer away from the interface. Simultaneously, the dynamics of the defined interface layer accurately reproduces those of related short-period systems. These observations regarding lattice dynamics significantly aid in determining the critical thickness of the interface.

The inter-atomic force constants (IFCs), from which we obtain the phonon frequencies and related dynamic matrices, are crucial physical quantities that can be transferred from first-principles calculations as key input parameters to determine model potentials. The reported results of IFCs in this study converge rapidly to the corresponding bulk behavior, aligned with the behavior of structural distortions and lattice dynamics, i.e., only one perovskite layer away from the interface. Furthermore, the transferability of the IFCs was preliminary confirmed on the phonon dispersions by artificially constructed IFCs matrix. Besides, as it is shown in the Sec. II. D of the SM , our proposal is compatible with the B site varies superlattices such as $(BaZrO_3)_n/(BaTiO_3)_m$. Based on these results, we can anticipate the development of a method for constructing transferable model potentials that properly capture the dynamics of layered oxide compounds with layers of arbitrary thicknesses.

## Acknowledgements

This work was financially supported by the National Natural Science Foundation of China (NSFC, Grant Nos.12274145), Guangdong Basic and Applied Basic Research Foundation, China (Grand No. 2023A1515010672). Guangdong Provincial University Science and Technology Program (Grant No. 2023KTSCX029). PhG was supported by the European Union's Horizon 2020 research and innovation program under Grant Agreement No. 964931 (TSAR) and by F.R.S.-FNRS Belgium under PDR Grant T.0107.20 (PROMOSPAN). Jin-Zhu Zhao appreciate the initial helpful discussions with Dr. Antoine Villesuzanne (from Institute of Condensed Matter Chemistry of Bordeaux, France) in the framework of IDS-FunMat PhD. Project. Jin-Zhu Zhao acknowledges the startup funding of South China Normal University.

# Supplemental Materials of
# Lattice dynamical properties and interatomic force constants of transition metal oxide perovskite superlattices

Xiao-Lin Chen,[1,2] Jin-Zhu Zhao,[1,2,3,4] Philippe Ghosez,[5]

*1. Guangdong Basic Research Center of Excellence for Structure and Fundamental Interactions of Matter, Guangdong Provincial Key Laboratory of Quantum Engineering and Quantum Materials, School of Physics, South China Normal University, Guangzhou 510006, P. R. China;*

*2. Guangdong-Hong Kong Joint Laboratory of Quantum Matter, Frontier Research Institute for Physics, South China Normal University, Guangzhou 510006, P. R. China;*

*3. Center for Computational Science and Engineering, Southern University of Science and Technology, Shenzhen 518055, P. R. China;*

*4. National Laboratory of Solid State Microstructures, Nanjing University, Nanjing 210093, P. R. China.*

*5. Theoretical Materials Physics, Q-MAT, University of Liège, B-4000 Sart Tilman, Belgium.*

## SECTION I: COMPUTATIONAL DETAILS

For cubic phase of bulk $ABO_3$, we observed that it is not a same level convergence when using the primitive cells to compare the results with $(PbTiO_3)_n/(SrTiO_3)_m$ and $(ATiO_3)_n/(AO)_m$ superlattices. The lattice constants of both $(PbTiO_3)_n/(SrTiO_3)_m$ and $(ATiO_3)_n/(AO)_m$ in the *c* direction is more than twice that of bulk $ABO_3$, which leads to certain differences in lattice vibration degrees of freedom between them. Therefore, for the cubic phase of bulk $ABO_3$, a *n*-layers supercell $(ABO_3)_n$ was used in order to compare the results at exactly the same level of convergence. For $(PbTiO_3)_1/(SrTiO_3)_3$ and $(PbTiO_3)_3/(SrTiO_3)_1$, we chose $(SrTiO_3)_4$ and tetragonal $(PbTiO_3)_4$ to calculate the phonon *DOS* and inter-atomic force-constants. For $(ATiO_3)_3/(AO)_2$ (AF phase, A=Ba、Sr and Pb) and $(ATiO_3)_3/(AO)_1$ (RP phase, A=Ba、Sr and Pb), we chose $(ATiO_3)_5$ and $(ATiO_3)_7$ to calculate the phonon *DOS* and inter-atomic force constants (*IFCs*) respectively.

## SECTION II: RESULTS AND DISCUSSIONS

### A. Phonon dispersion curves

We show phonon dispersion curves of $(BaTiO_3)_n/(BaO)_m$ (n=1,2; m=1,2) in Fig. S1. We observe that the high-symmetry phase of these superlattices show unstable polar phonon mode at Γ point. The eigenvectors of the these modes correspond to the in-plane polar motion of the ions. Further atomic relaxations confirm that only in-plane polarization appears while the out-of-plane component of polarization is suppressed, we report the result in Table S1. These results agree with the previous understanding of the emergence of ferroelectricity, thus, the out-of-plane polarization disappears due to the long-range correlation along *z* direction is broken by the interface. In the meantime, the in-plane polarization survives since the in-plane periodicity remains.

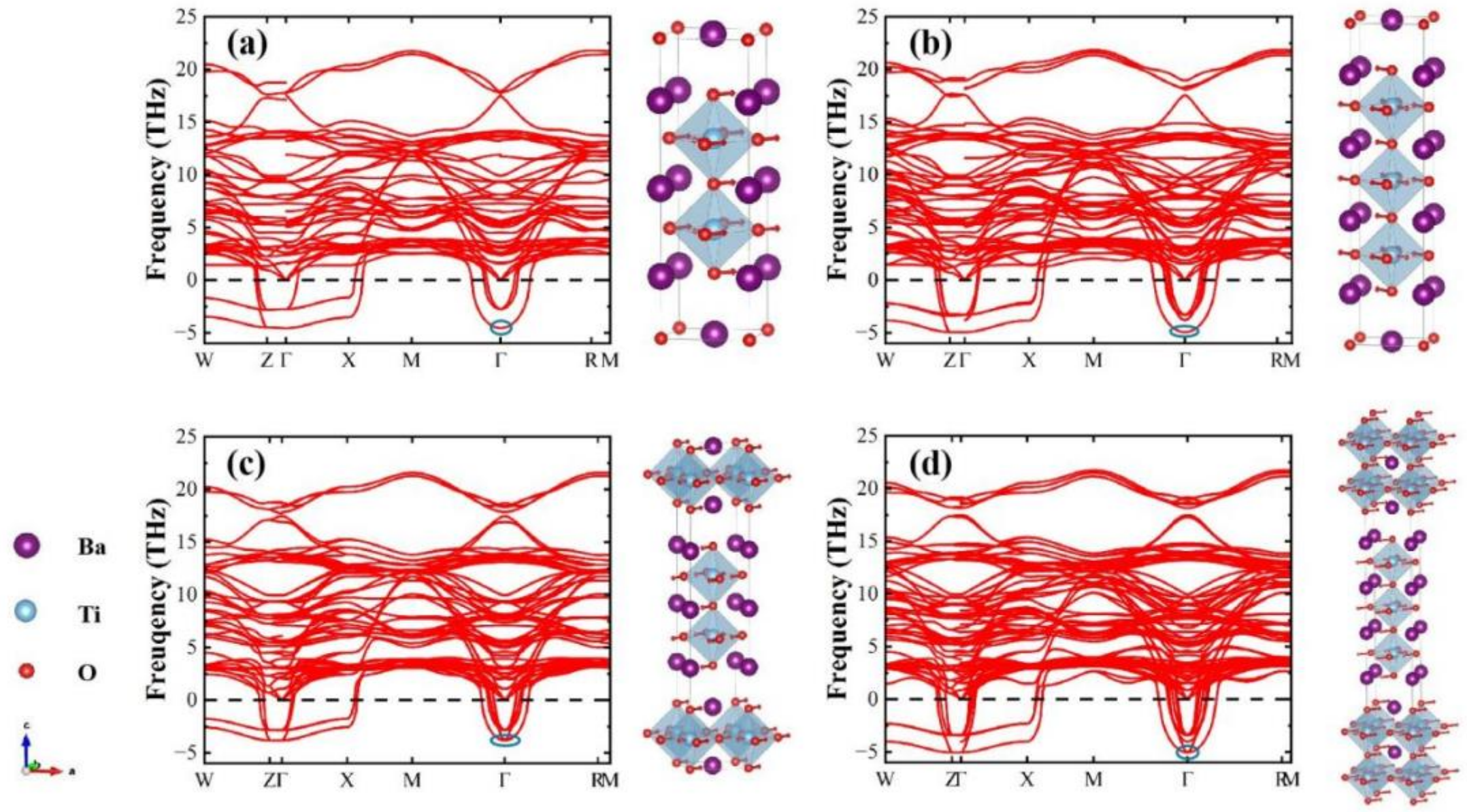

**Figure S1.** Phonon dispersion curves and the atomic motion along the eigenvectors of the lowest mode at Γ point in $(BaTiO_3)_n/(BaO)_m$ (n=1,2; m=1,2) superlattices. (a) n=2, m=2; (b) n=3, m=2; (c) n=2, m=1; (d) n=3, m=1.

**Table S1.** The components of polarization（$\mu C/cm^2$）in ground states of $(BaTiO_3)_m/(BaO)_n$ (n=2,3; m=1,2) superlattices.

| | $p_x$ | $p_y$ | $p_z$ |
|---|---|---|---|
| n=2,m=2 | -9.5 | -10.0 | 0 |
| n=3,m=2 | 8.5 | -9.3 | 0 |
| n=2,m=1 | 0.4 | -1.0 | 0 |
| n=3,m=1 | -0.1 | -3.3 | 0 |

## B. $(ATiO_3)_n/(AO)_m$ (A=Sr and Pb)

### *1. Structural convergence*

In the same way, the $(ATiO_3)_n/(AO)_m$ (A=Sr and Pb) superlattices are relaxed without any epitaxial stain. The results of Sr based systems and Pb based systems are reported in Table S2 and Table S3 respectively. The conclusion is consistent with the $(PTO)_n/(STO)_m$ and Ba based systems. The lattice distortions are mainly appeared in the interface region and could be ignored in the central region. Whether AF phase or RP phase of Sr based systems and Pb based systems, the thicknesses of the central layer in large-period superlattices ($n$=3) is close to value of bulk $ATiO_3$ (A= Sr, Pb). These values of interface layer also nicely recover the values in corresponding short-period superlattices.

**Table S2.** In-plane lattice constants (a), the thicknesses of interface perovskite layer ($c_i$) and central perovskite layer ($c_c$), rumpling in the interface perovskite layer, which is defined as the displacement along $z$ between cations and oxygen atoms $R_{rump}=(z_{cat}-z_O)/2$, Ti-O bond length, under zero epitaxial strain of Sr based superlattices for AF and RP types (all distances in Å)

| | | Lattice size | | | Atomic rumpling | | | Ti-O bond length | | |
|---|---|---|---|---|---|---|---|---|---|---|
| | | a | $c_i$ | $c_c$ | layer 1 | layer 2 | layer 3 | bond 1 | bond 2 | bond 3 |
| AF | n=1 | 3.830 | 3.865 | — | 0.063 | — | — | 1.995 | — | — |
| | n=2 | 3.860 | 3.872 | — | 0.074 | 0.011 | — | 1.960 | 1.986 | — |
| | n=3 | 3.872 | 3.873 | 3.896 | 0.076 | 0.013 | 0.011 | 1.951 | 1.988 | 1.959 |
| RP | n=1 | 3.879 | 3.806 | — | 0.083 | — | — | 1.986 | — | — |
| | n=2 | 3.893 | 3.835 | — | 0.098 | 0.016 | — | 1.952 | 1.981 | — |
| | n=3 | 3.896 | 3.844 | 3.874 | 0.102 | 0.018 | 0014 | 1.946 | 1.985 | 1.952 |
| $SrTiO_3$ | | | 3.899 | | | 0 | | | 1.950 | — |

**Table S3.** In-plane lattice constants (a), the thicknesses of interface perovskite layer ($c_i$) and central perovskite layer ($c_c$), rumpling in the interface perovskite layer, which is defined as the displacement along $z$ between cations and oxygen atoms $R_{rump}=(z_{cat}-z_O)/2$, Ti-O bond length, under zero epitaxial strain of Pb based superlattices for AF and RP types (all distances in Å)

| | | Lattice size | | | Atomic rumpling | | | Ti-O bond length | | |
|---|---|---|---|---|---|---|---|---|---|---|
| | | a | $c_i$ | $c_c$ | layer 1 | layer 2 | layer 3 | bond 1 | bond 2 | bond 3 |
| AF | n=1 | 3.846 | 3.965 | — | 0.026 | — | — | 2.007 | — | — |
| | n=2 | 3.882 | 3.938 | — | 0.035 | 0.006 | — | 1.972 | 2.001 | — |
| | n=3 | 3.897 | 3.929 | 3.930 | 0.038 | 0.009 | 0.007 | 1.959 | 2.001 | 1.972 |
| RP | n=1 | 3.891 | 3.906 | — | 0.041 | — | — | 1.994 | — | — |
| | n=2 | 3.911 | 3.899 | — | 0.054 | 0.010 | — | 1.956 | 1.997 | — |
| | n=3 | 3.920 | 3.897 | 3.899 | 0.060 | 0.014 | 0.012 | 1.946 | 1.985 | 1.952 |
| $PbTiO_3$ | | | 3.924 | | | 0 | | | 1.962 | — |

*2. Phonon Density of States*

As shown in Fig. S2 and Fig. S3, we report the phonon density of states of the Sr based superlattices and Pb based superlattices. Obviously, whether $(SrTiO_3)_n/(SrO)_m$ or $(PbTiO_3)_n/(PbO)_m$ superlattices, the phonon *DOS* of AF structures and RP structures in central layer is different with the ones in interface layer. Fig. S2(a) shows a peak in the negative frequency range between -3 THz to 0 THz, corresponds to the ferroelectric instability and unstable rotation mode. In Fig. S2(c), the peak weak is a negligible value. In the frequency range from 0 to 5 THz, there mainly is a peak, which corresponds to the contribution of Sr atoms from the central pervoskite layer. We found that RP compounds have similarly results from Fig.S2 (b) and (d).

In the negative frequency range, there is a peak around -1.0 THz, which corresponds to the ferroelectric instability, as it is shown in Fig. S3 (a). However, in Fig. S3 (c), the amplitude of instability in the interface perovskite layer is larger than the central perovskite layer, which corresponds to an enhanced ferroelectric instability. In the positive frequency region, there are two large amplitude peaks in the DOS of cubic $PbTiO_3$. The first one, around 1.5 THz, corresponds to the contribution of Pb atoms. The second one, corresponds to the contribution of $TiO_2$ layers, is found at around 14.3 THz. In Fig. S3 (b) and (d), the RP types have similar results.

These results confirm that whether Sr based systems or Pb based systems, the phonon *DOS* of the AF phase and RP phase in the central layer recover the bulk $ABO_3$ values. Meanwhile, the phonon *DOS* of the interface layer recover the related short-period superlattices.

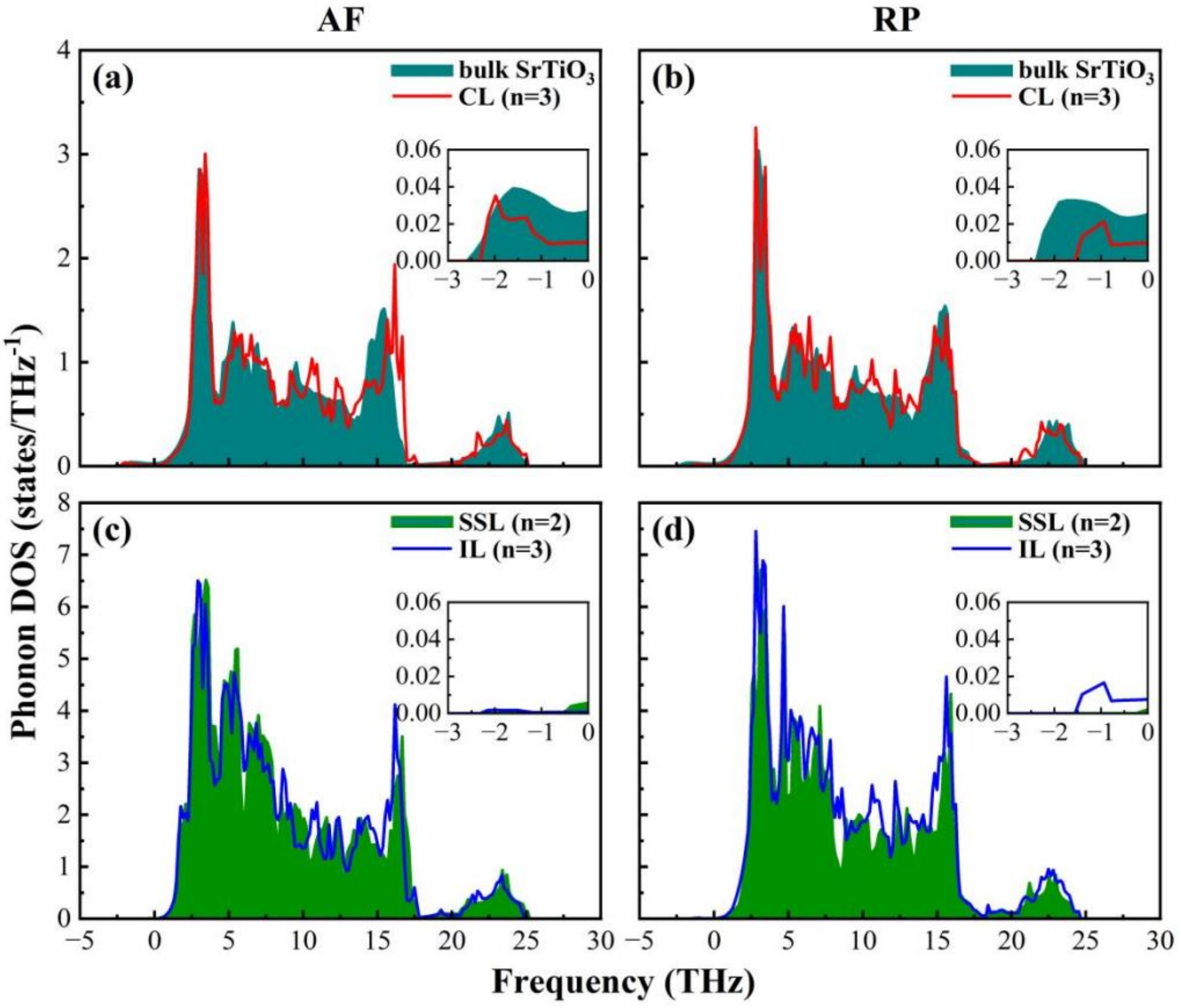

**Figure S2.** Phonon density of states (*DOS*) of Sr based superlattices in the case of AF and RP types. The region of negative frequencies is zoomed as the insets. (a) and (b), the dark cyan color part refers the phonon *DOS* of the bulk $SrTiO_3$, the red line indicates the phonon *DOS* contribution of the central perovskite layer in *n*=3 superlattices; (c) and (d), the blue line indicates the phonon *DOS* contribution of the interface perovskite layer in *n*=3 superlattices, the dark green color refers the contribution from the short-period superlattices in case of *n*=2.

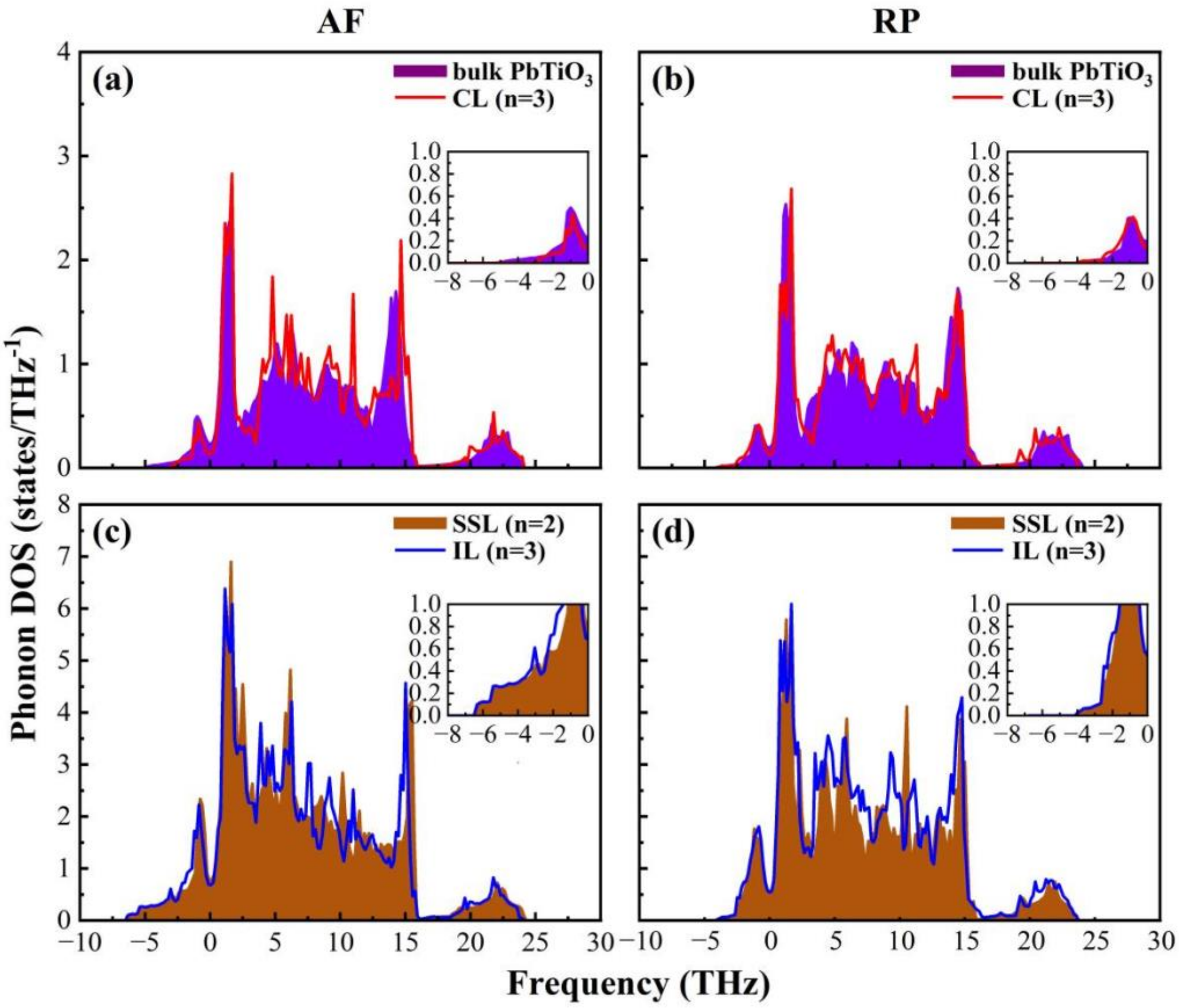

**Figure S3.** Phonon density of states (*DOS*) of Pb based superlattices in the case of AF and RP types. The region of negative frequencies is zoomed as the inset. (a) and (b), the purple color part refers the phonon *DOS* of the bulk $PbTiO_3$, the red line indicates the phonon *DOS* contribution of the central perovskite layer in *n*=3 superlattices; (c) and (d), the blue line indicates the phonon *DOS* contribution of the interface perovskite layer in *n*=3 superlattices, the brown color refers the contribution from the short-period superlattices in case of *n*=2.

*3. Inter-atomic Force Constants*

We report the *IFCs* values affecting on different Ti atoms which are selected in central perovskite layer and interface perovskite layer of Sr and Pb based superlattices, including AF and RP phase(*n*=3). Furthermore, the difference of *IFCs* values between these Ti atoms and the ones in the related bulk case and short-period systems (*n*=2) are illustrated as well. All results are reported in Fig. S4 - 7 and similar with Ba based systems. As it is illustrated in these figures, the absolute *IFCs* values of Ti atom are very similar between the case in AF and RP superlattices of Sr and Pb based superlattices. We found that the *IFCs* values of AF and RP structures in the central perovskite layer recover the values of bulk $ABO_3$. The *IFCs* values of the interface pervoskite layer recover the values of related short- period supperlattices.

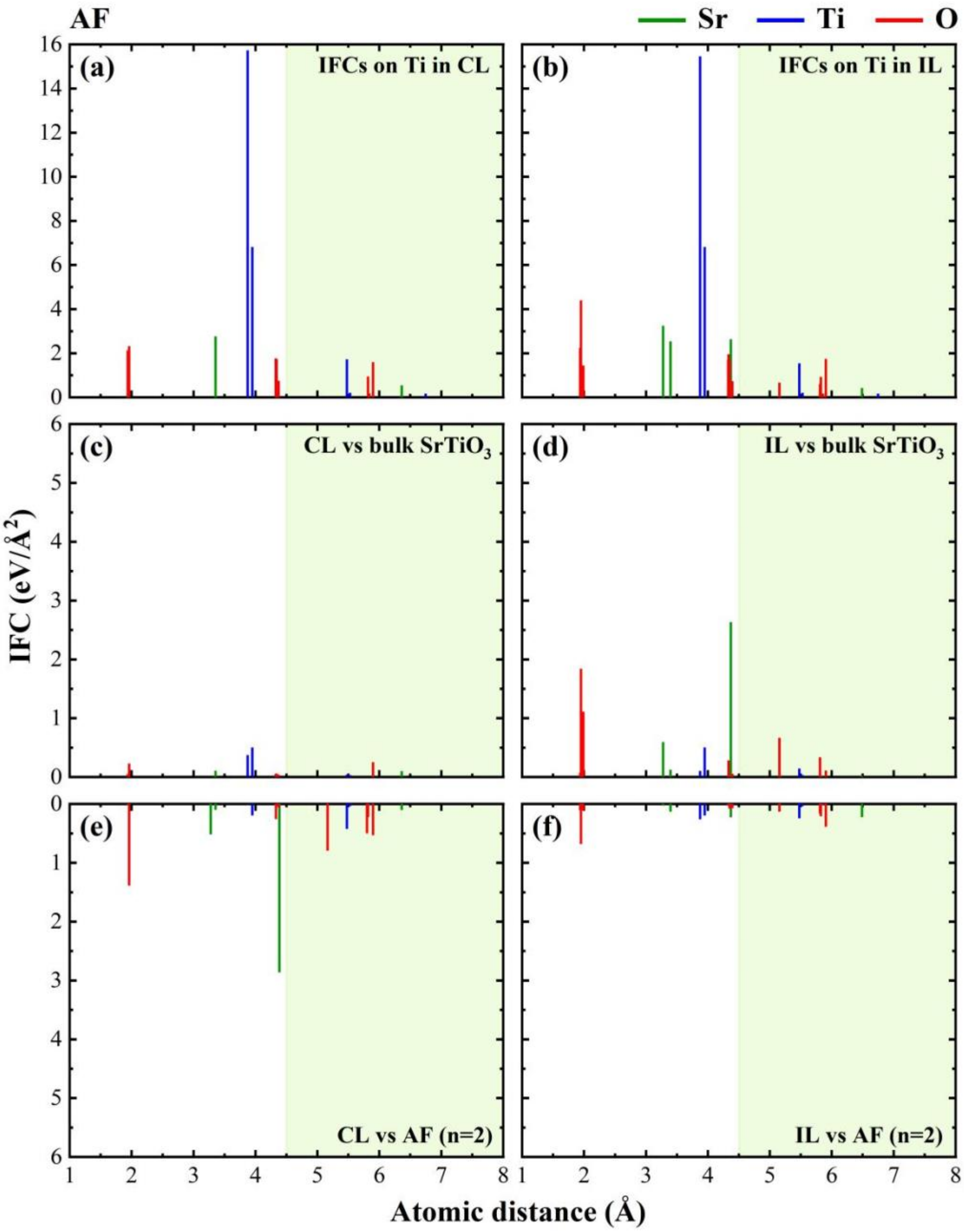

**Figure S4.** Absolute difference between values of inter-atomic constants (*IFCs*) affect on Ti atoms in different selected systems: (a) the central layer of AF superlattice (*n*=3), (b) the interface layer of AF superlattice (*n*=3). Difference between *IFCs*' for Ti atom (c) in bulk $SrTiO_3$ and the ones in central layer of AF superlattice (*n*=3); (d) in bulk $SrTiO_3$ and the ones in interface layer of AF superlattice (*n*=3); (e) in short-period AF superlattice (*n*=2) and the ones in central layer AF superlattice (*n*=3); (f) in short-period AF superlattice (*n*=2) and the ones in interface layer AF superlattice (*n*=3).

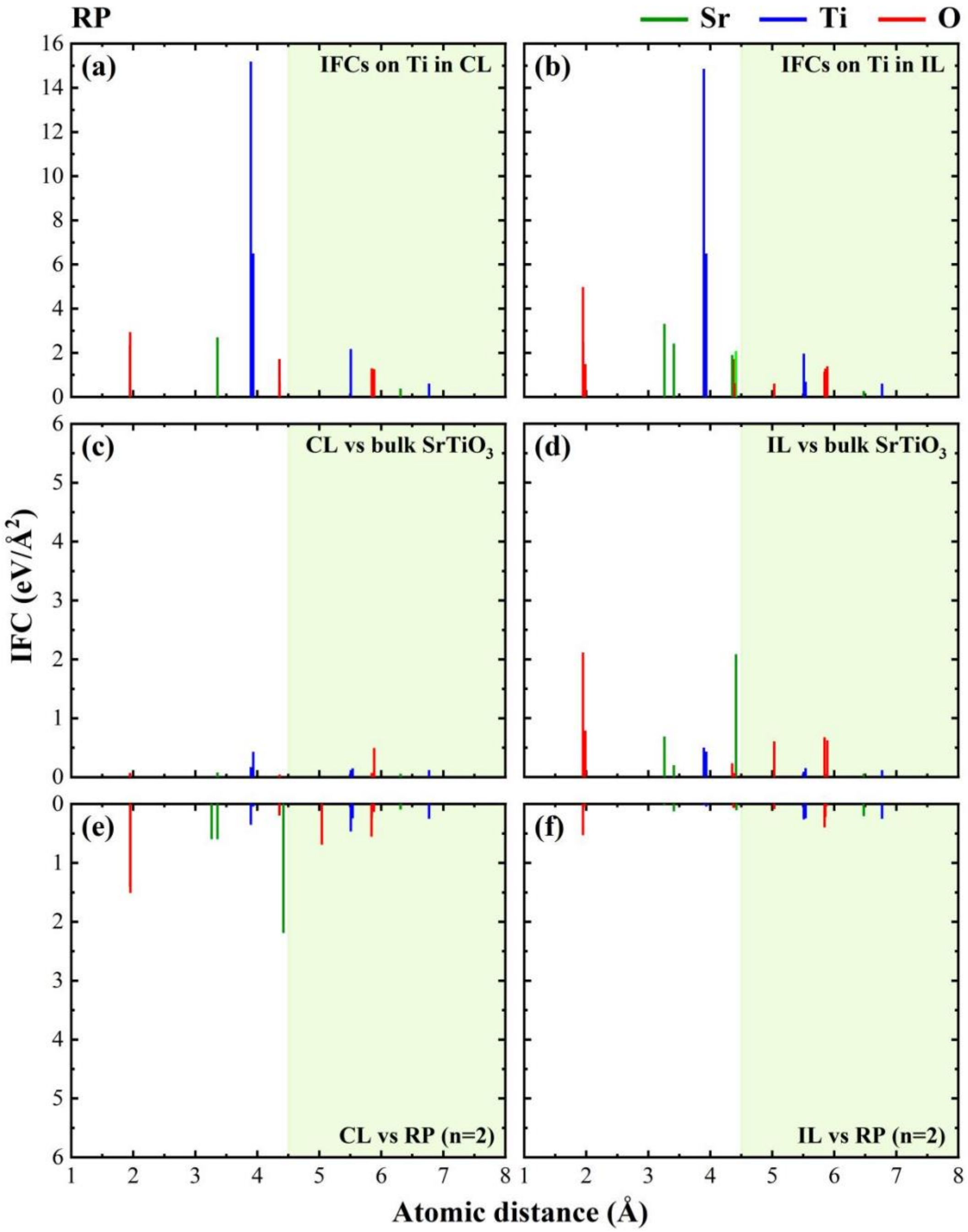

**Figure S5.** Absolute difference between values of inter-atomic constants (*IFCs*) affect on Ti atoms in different selected systems: (a) the central layer of RP superlattice (*n*=3), (b) the interface layer of RP superlattice (*n*=3). Difference between *IFCs*' for Ti atom (c) in bulk SrTiO3 and the ones in central layer of RP superlattice (*n*=3); (d) in bulk $SrTiO_3$ and the ones in interface layer of RP superlattice (*n*=3); (e) in short-period RP superlattice (*n*=2) and the ones in central layer RP superlattice (*n*=3); (f) in short-period RP superlattice (*n*=2) and the ones in interface layer RP superlattice (*n*=3).

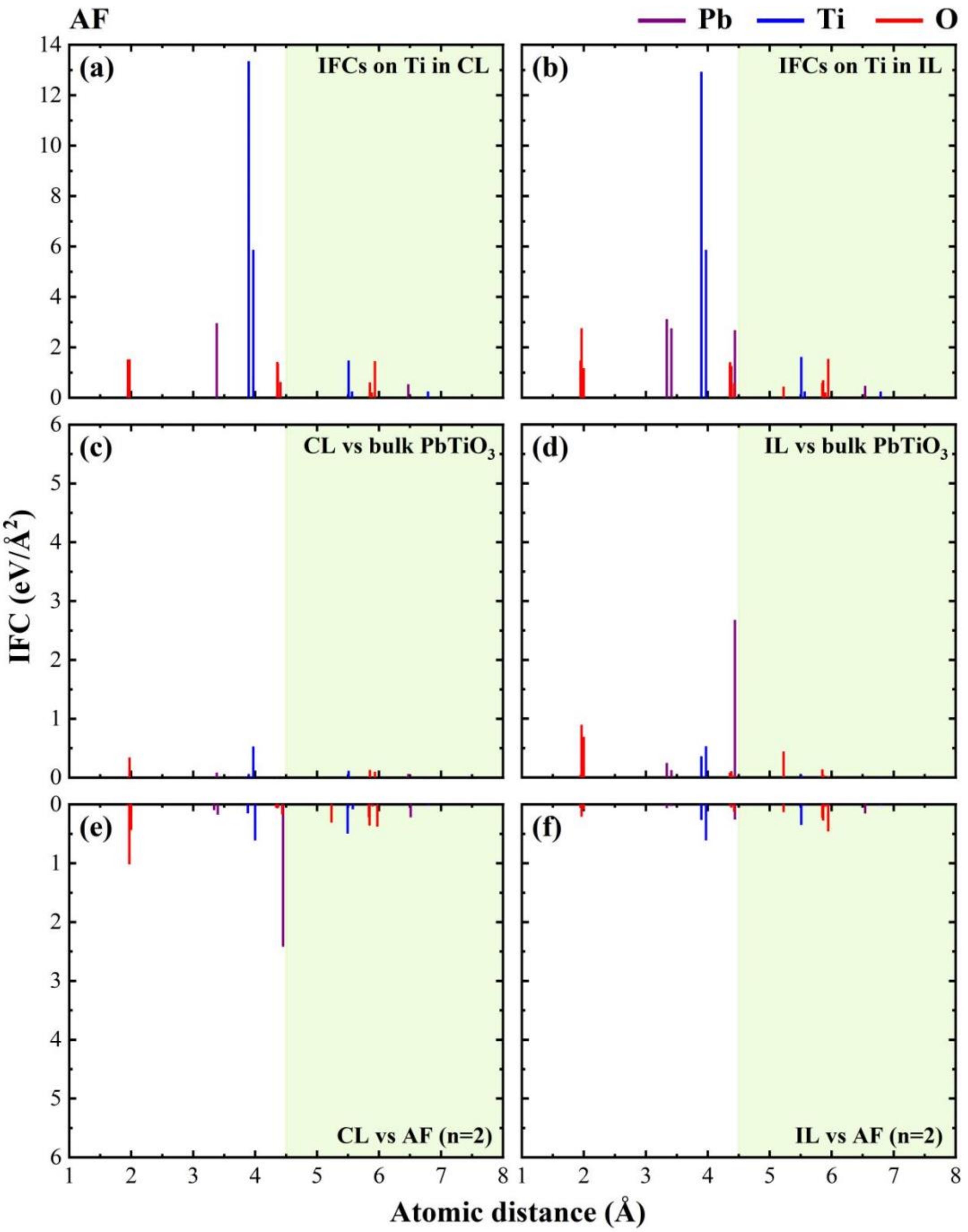

**Figure S6.** Absolute difference between values of inter-atomic constants (*IFCs*) affect on Ti atoms in different selected systems: (a) the central layer of AF superlattice (*n*=3), (b) the interface layer of AF superlattice (*n*=3). Difference between *IFCs*' for Ti atom (c) in bulk $PbTiO_3$ and the ones in central layer of AF superlattice (*n*=3); (d) in bulk $PbTiO_3$ and the ones in interface layer of AF superlattice (*n*=3); (e) in short-period AF superlattice (*n*=2) and the ones in central layer AF superlattice (*n*=3); (f) in short-period AF superlattice (*n*=2) and the ones in interface layer AF superlattice (*n*=3).

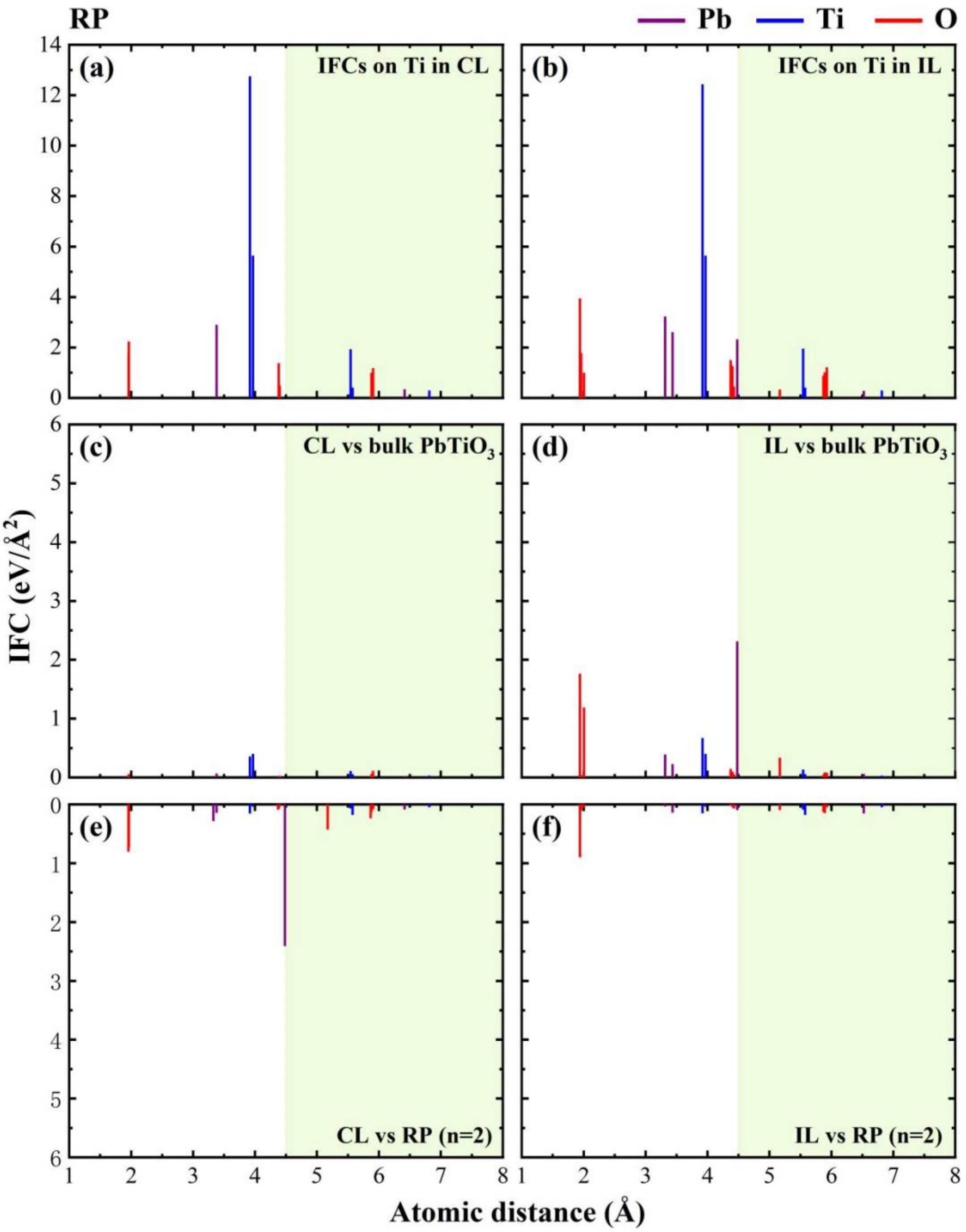

**Figure S7.** Absolute difference between values of inter-atomic constants (*IFCs*) affect on Ti atoms in different selected systems: (a) the central layer of RP superlattice (*n*=3), (b) the interface layer of RP superlattice (*n*=3). Difference between *IFCs*' for Ti atom (c) in bulk PbTiO3 and the ones in central layer of RP superlattice (*n*=3); (d) in bulk $PbTiO_3$ and the ones in interface layer of RP superlattice (*n*=3); (e) in short-period RP superlattice (*n*=2) and the ones in central layer RP superlattice (*n*=3); (f) in short-period RP superlattice (*n*=2) and the ones in interface layer RP superlattice (*n*=3).

## C. *The truncation value for IFCs*

In the manuscript we proposed that the spatial truncation can be chosen in one perovskite layer. The relating minimum absolute values of the *IFCs* are 0.1 eV/Å$^2$. We further test this analysis on phonon behaviors of studied compounds. In our additional calculations, the *IFCs* are artificially suppressed when the value of which is below 0.1 eV/Å$^2$. The phonon dispersion curves and *DOS* are nicely reproduced via these artificially constructed (A.C.) *IFCs* matrix, as shown in Fig. S8, indicating that the contribution of *IFCs* on lattice dynamics can be ignored when it is small enough. Not only for bulk $SrTiO_3$ and $PbTiO_3$, this argument is compatible with the case of tested superlattices P1S3.

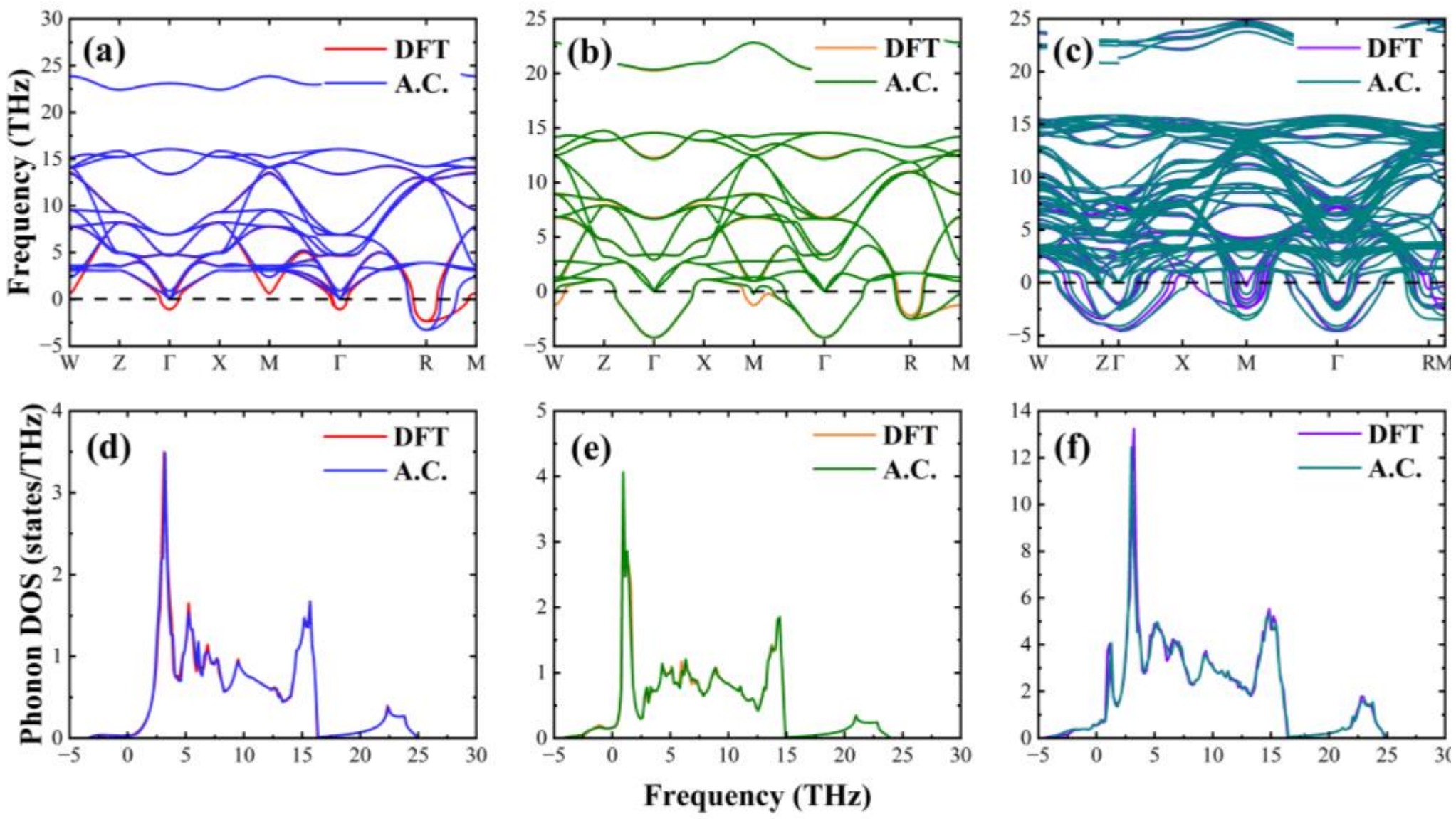

**Figure S8.** The comparison between phonon dispersion curves and phonon density of states obtained from the original DFT data and those obtained from the data of *IFCs* below 0.1 eV/Å$^2$ is ignored (artificially constructed, A.C.). (a) phonon dispersion curves and (d) phonon density of states obtained from the original DFT data (red lines) and the data (blue lines) of *IFCs* below 0.1 eV/Å$^2$ is ignored for bulk $PbTiO_3$; (b) phonon dispersion curves and (e) phonon density of states obtained from the original DFT data (orange lines) and the data (light green lines) of *IFCs* below 0.1 eV/Å$^2$ is ignored for bulk $SrTiO_3$; (c) phonon dispersion curves and (f) phonon density of states obtained from the original DFT data (purple lines) and the data (cyan lines) of *IFCs* below 0.1 eV/Å$^2$ is ignored for P1S3 superlattice.

**D. $(BaZrO_3)_n/(BaTiO_3)_m$**

Here we report the results on $(BaZrO_3)_n/(BaTiO_3)_m$ superlattices. In our calculations, the in-plane lattice constants are fixed to the cubic $BaZrO_3$ values, and the space group of the superlattices is constrained to the high-symmetry *P4/mmm* phase. We allow the *c*-axis (along the *z*-direction) and the internal atomic degrees of freedom to relax together. In Table S4, we report the values characterizing the structural distortions. It is shown that the values in the interface region of large-period superlattices (n, m=2 or 3) are close to those of the corresponding short-period superlattice (n, m=1), while the values in the central region almost recover those of the bulk materials.

We obtain the same conclusion as the main text regarding the phonon density of states in reciprocal space and the inter-atomic force constants (*IFCs*) in real space. As shown in Figure S9, S10 and S11, the phonon density of states and *IFCs* in the central layer of the large-period superlattices are well restored to the values of the corresponding bulk materials, while those in the interface layer closely recover the values of the short-period superlattices. This demonstrates that our conclusion is applicable for the B site varies superlattices.

**Table S4.** Lattice size and thicknesses of central layer $c_c$ and interface layer $c_i$ (in Å) in $(BZO)_n/(BTO)_m$ (n, m=2 or 3), compared with related values in bulk cubic $BaZrO_3$ and bulk cubic $BaTiO_3$. Value of rumpling for each layer (in Å) in $(BZO)_n/(BTO)_m$. The length of Ti-O bond along *z* in $(BZO)_n/(BTO)_m$ compared with related values in bulk cubic $BaZrO_3$ and bulk cubic $BaTiO_3$.

| | Lattice size | | Atomic rumpling | | | | | | Ti-O bond length | | | | |
|---|---|---|---|---|---|---|---|---|---|---|---|---|---|
| | $c_i$ | $c_c$ | layer 1 | layer 2 | layer 3 | layer 4 | layer 5 | Layer 6 | bond 1 | bond 2 | bond 3 | bond 4 | bond 5 |
| $(BZO)_2/(BTO)_3$ | 8.059 | 3.984 | 0 | 0.002 | -0.003 | 0.001 | 0.001 | 0 | 2.085 | 2.117 | 1.907 | 1.945 | 1.992 |
| $(BZO)_3/(BTO)_2$ | 8.053 | 4.192 | 0 | -0.001 | 0.002 | -0.002 | -0.001 | 0 | 1.942 | 1.910 | 2.117 | 2.084 | 2.095 |
| BZOBTO | 8.055 | | 0 | 0.004 | 0.001 | -0.002 | — | — | 1.940 | 1.911 | 2.114 | 2.090 | — |
| $BaZrO_3$ | 4.197 | | 0 | | — | — | — | — | 2.099 | | — | — | — |
| $BaTiO_3$ | 3.988 | | 0 | | — | — | — | — | 1.994 | | — | — | — |

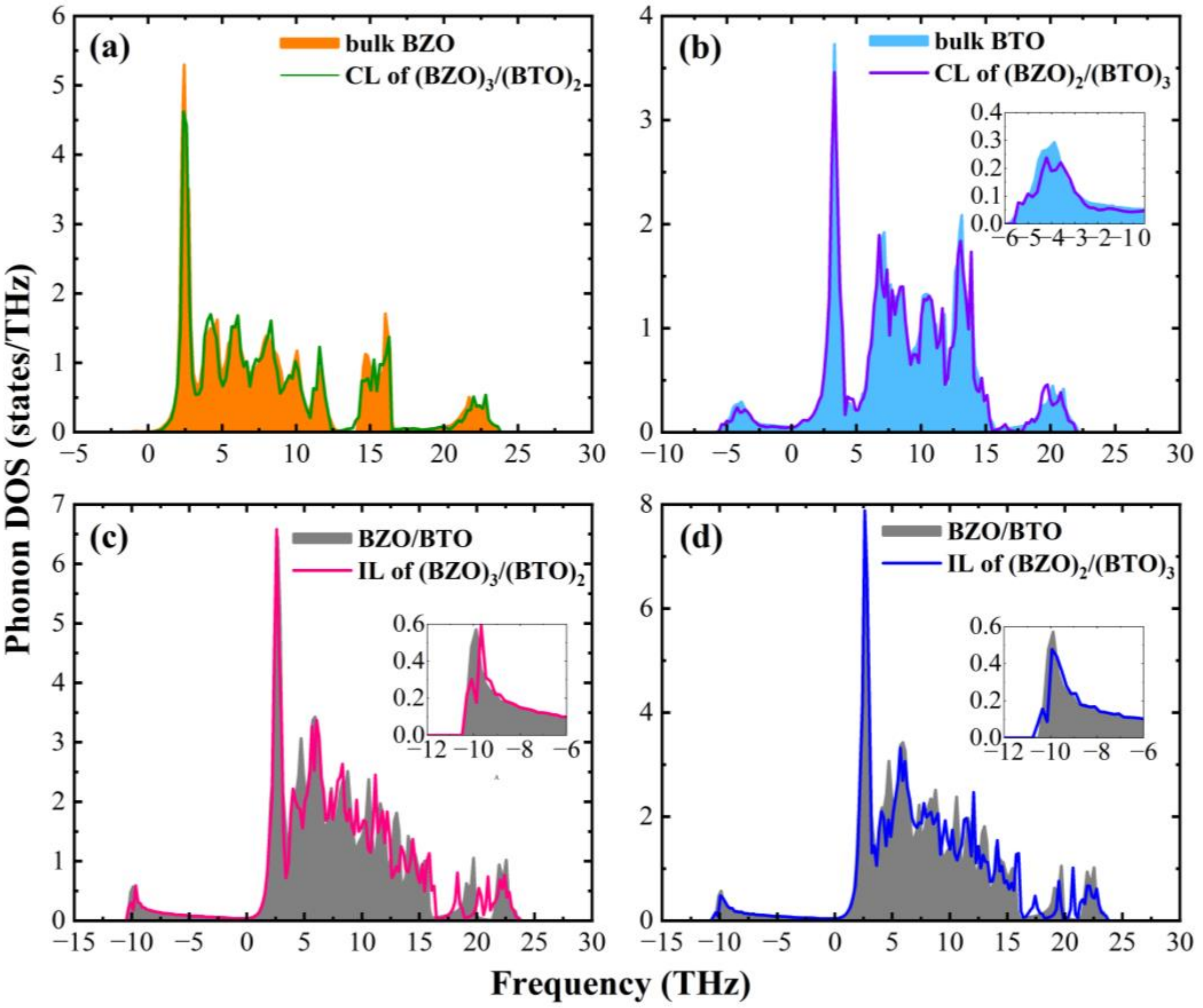

**Figure S9.** Phonon density of states (DOS) of interface and central layers in $(BZO)_2/(BTO)_3$ and $(BZO)_3/(BTO)_2$ cases, and the phonon DOS for bulk. The region of negative frequencies is zoomed as insets. (a) bulk cubic $BaZrO_3$ (orange line) and central layer in $(BZO)_3/(BTO)_2$ (green line); (b) bulk cubic $BaTiO_3$ (light blue line) and central layer in $(BZO)_2/(BTO)_3$ (purple line); (c) BZO/BTO case (gray line) and interface layer in $(BZO)_3/(BTO)_2$ (pink line); (d) BZO/BTO case (gray line) and interface layer in $(BZO)_2/(BTO)_3$ (blue line)

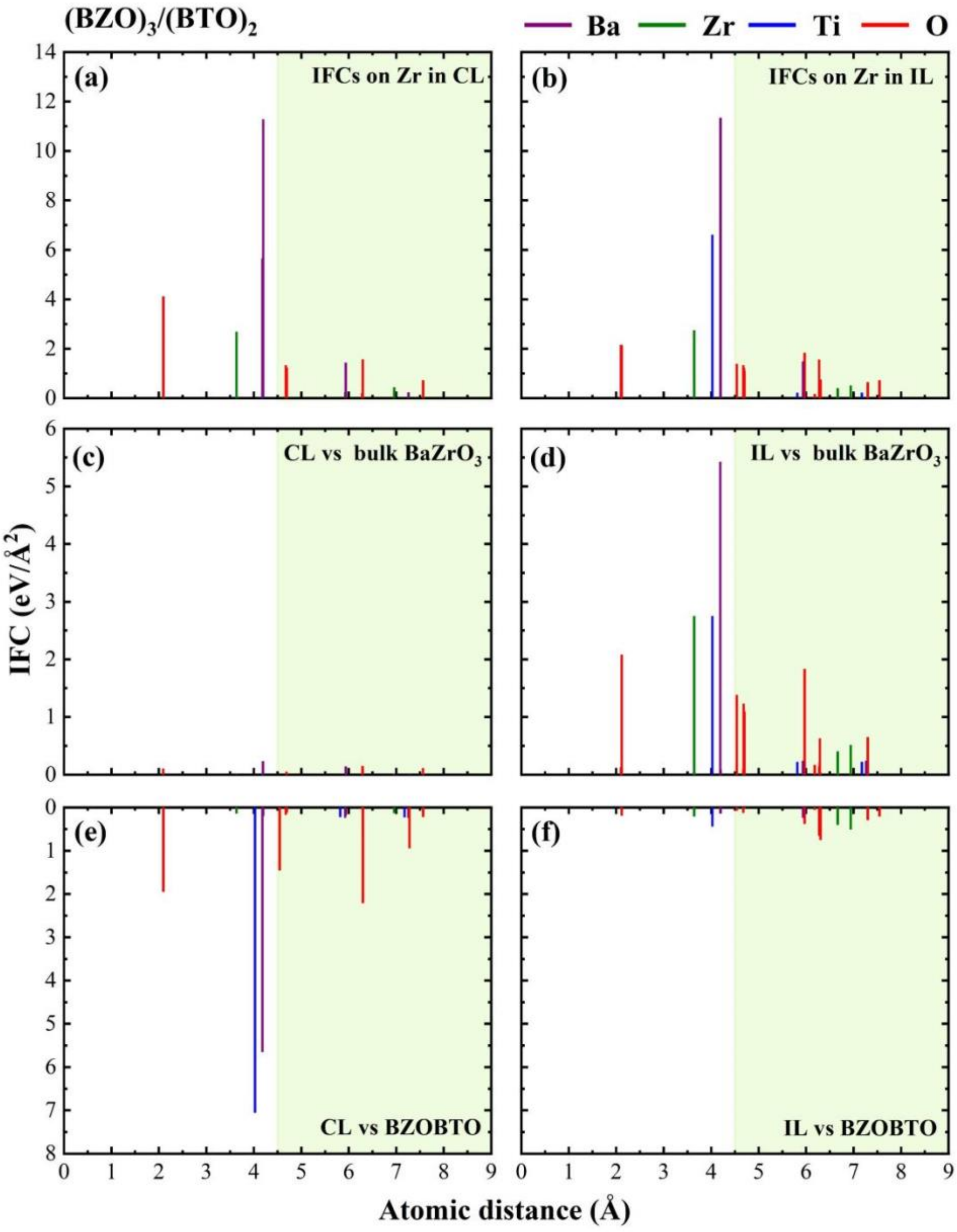

**Figure S10.** Absolute value of *IFCs* difference affect on Ti atoms in difference selected systems: (a) the central layer of $(BZO)_3/(BTO)_2$, (b) the interface layer of $(BZO)_3/(BTO)_2$. And the absolute difference between values of *IFCs* affect on Ti atoms in different selected systems: the difference between *IFCs*' for Ti atom. (c) in bulk cubic $BaZrO_3$ and the ones in central layer of $(BZO)_3/(BTO)_2$; (d) in tetragonal bulk $PbTiO_3$ and the ones in interface layer of $(BZO)_3/(BTO)_2$; (e) in SSL BZO/BTO and the ones in central layer of $(BZO)_3/(BTO)_2$; (f) in SSL BZO/BTO and the ones in interface layer of $(BZO)_3/(BTO)_2$.

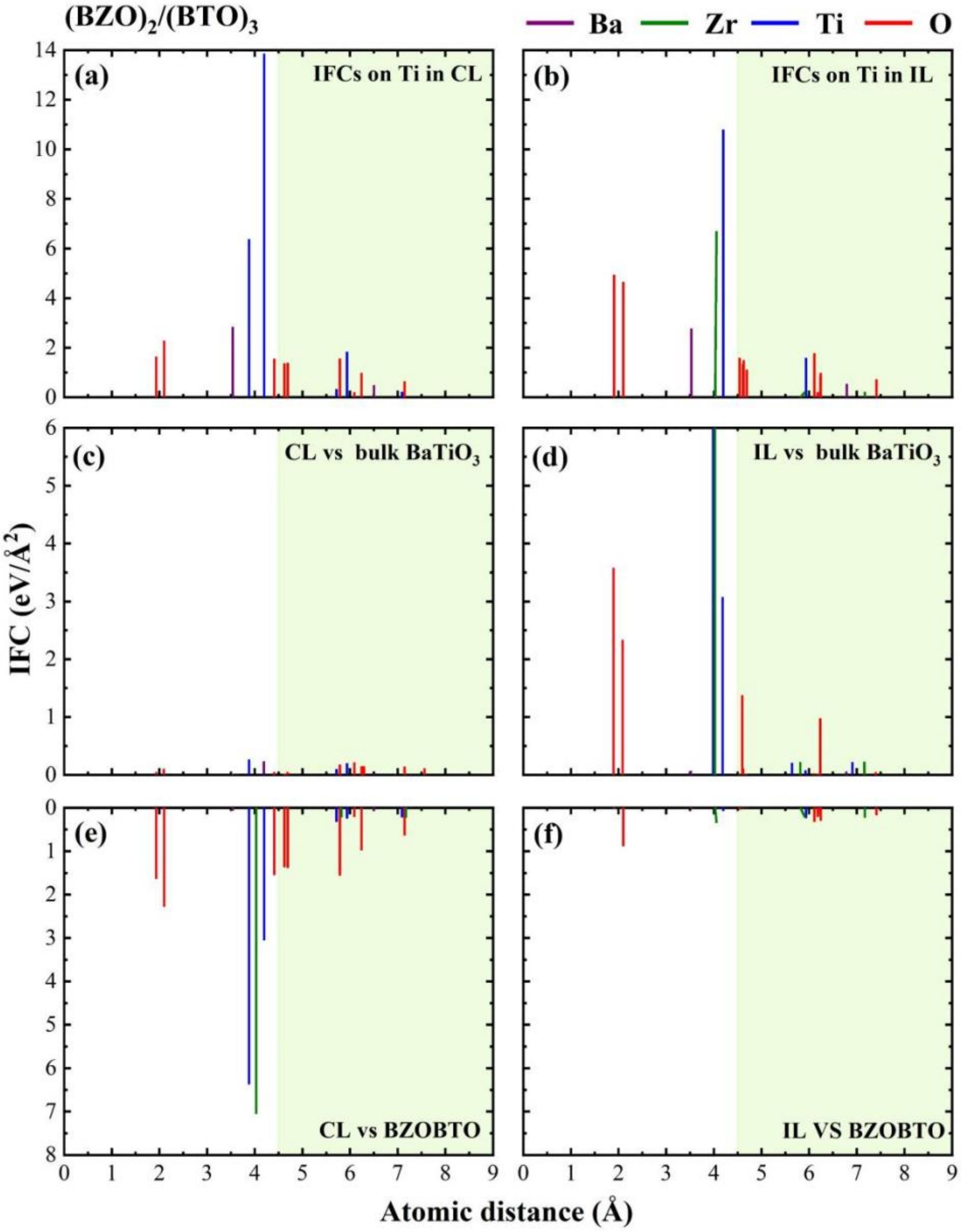

**Figure S11.** Absolute value of *IFCs* difference affect on Ti atoms in difference selected systems: (a) the central layer of $(BZO)_2/(BTO)_3$, (b) the interface layer of $(BZO)_2/(BTO)_3$. And the absolute difference between values of *IFCs* affect on Ti atoms in different selected systems: the difference between *IFCs*' for Ti atom. (c) in bulk cubic $BaTiO_3$ and the ones in central layer of $(BZO)_2/(BTO)_3$; (d) in bulk cubic $PbTiO_3$ and the ones in interface layer of $(BZO)_2/(BTO)_3$; (e) in SSL BZO/BTO and the ones in central layer of $(BZO)_2/(BTO)_3$; (f) in SSL BZO/BTO and the ones in interface layer of $(BZO)_2/(BTO)_3$.